\documentclass[compsoc, conference, letterpaper, 10pt, times]{IEEEtran}

\usepackage{amsmath,amsthm,amsfonts}
\usepackage{hyperref}
\usepackage{tikz}
\usepackage{balance}
\ifCLASSOPTIONcompsoc
  \usepackage[nocompress]{cite}
\else
  \usepackage{cite}
\fi
\ifCLASSINFOpdf
\else
\fi
\begin{document}
%
\title{Quantifying the Information Leak in Cache Attacks through Symbolic Execution}

\author{\IEEEauthorblockN{Sudipta Chattopadhyay}
\IEEEauthorblockA{Saarland University}
\and
\IEEEauthorblockN{Moritz Beck}
\IEEEauthorblockA{Saarland University}
\and
\IEEEauthorblockN{Ahmed Rezine}
\IEEEauthorblockA{Link\"oping University}
\and
\IEEEauthorblockN{Andreas Zeller}
\IEEEauthorblockA{
Saarland University}}



\clubpenalty=10000
\widowpenalty=10000

\newtheorem{theorem}{Theorem}[section]
\newtheorem{definition}[theorem]{Definition}

\newcommand{\todo}[1]{{\color{red} TODO: #1}}%

\maketitle

\begin{abstract}
Cache timing attacks allow attackers to infer the 
properties of a secret execution by observing cache 
hits and misses. But how much information can actually 
leak through such attacks? For a given program, a cache 
model, and an input, our CHALICE framework leverages 
symbolic execution to compute the amount of information 
that can possibly leak through cache attacks. At the core 
of CHALICE is a novel approach to quantify information 
leak that can highlight critical cache side-channel 
leaks on arbitrary binary code. In our evaluation on  
real-world programs from OpenSSL and Linux GDK libraries, 
CHALICE effectively quantifies information leaks: For an  
AES-128 implementation on Linux, for instance, CHALICE 
finds that a cache attack can leak as much as 127 out 
of 128 bits of the encryption key.

\end{abstract}


%
\IEEEpeerreviewmaketitle

\section{Introduction}

Cache timing attacks~\cite{djb:2005cache} are among the best known \emph{side channel} attacks to determine secret features of a program execution without knowing its input or output.  The general idea of a timing attack is to observe, for a known program, a timing of cache hits and misses, and then to use this timing to determine or constrain features of the program execution, including secret data that is being processed.

The precise nature of the information that \emph{can} leak through such attacks depends on the cache and its features, as well as the program and its features. Consequently, given a model of the cache and a program run, it is possible to analyze which and how much information would leak through a cache attack.  This is what we do in this paper.  Given a program execution and a cache model, our CHALICE approach automatically determines \emph{which bits of the input would actually leak through a potential cache attack.}

As an example, consider an implementation of the popular AES encryption algorithm.  Given an input and an encryption key (say, 128 bits for AES-128), CHALICE can determine which and how many of the bits of the key would leak if the execution were subject to a cache attack.  To this end, CHALICE uses a novel \emph{symbolic execution} over the given concrete input. 
During symbolic execution, CHALICE derives symbolic timings of cache hits and misses; these then again reveal under which circumstances individual bits of encryption key may leak.

\begin{figure}[t]
\begin{center}
\rotatebox{0}{
\includegraphics[scale = 0.27]{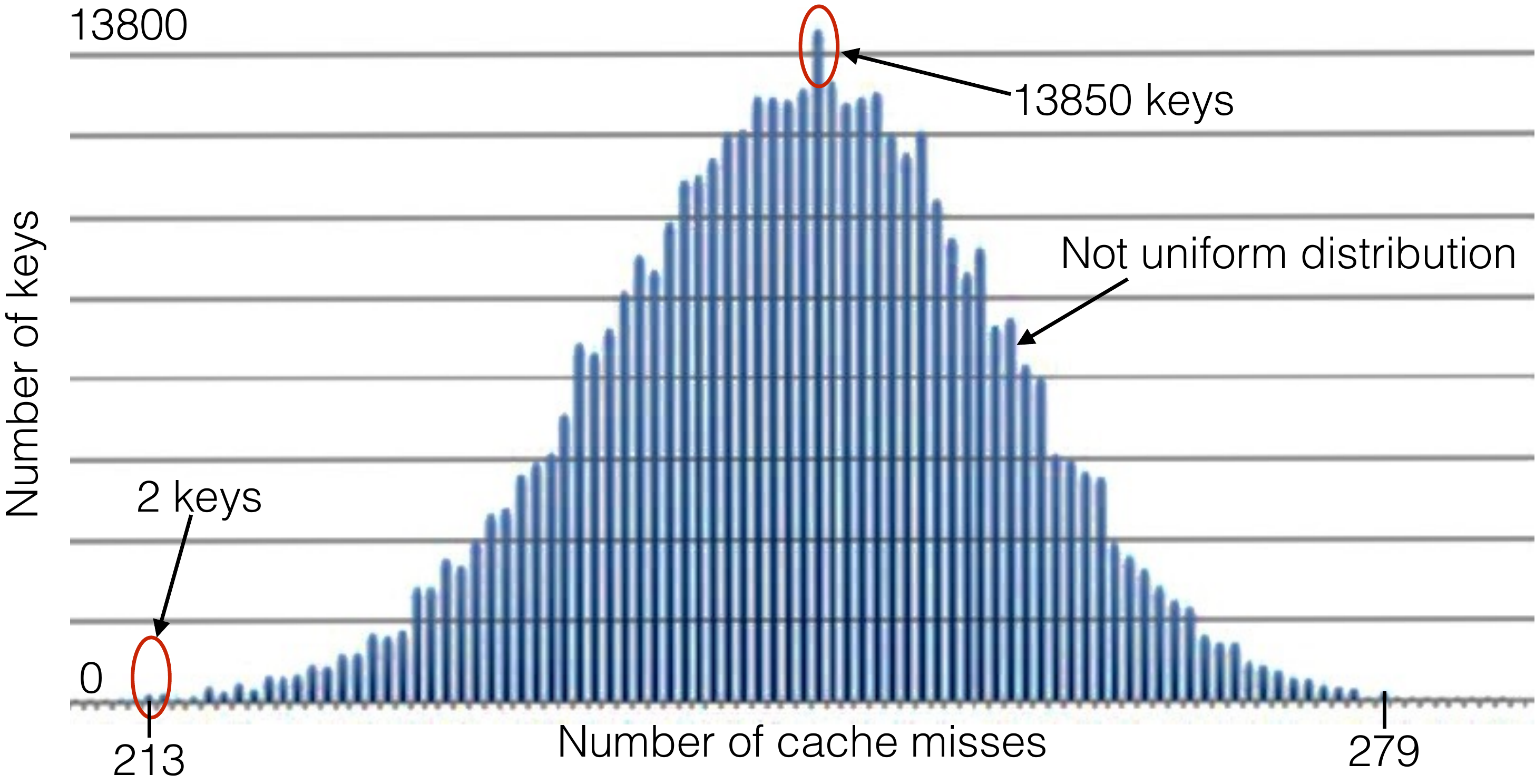}}
\vspace*{-0.1in}
\caption{For a fixed input message, the plot shows the number of keys 
leading to a given number of cache misses incurred by executing AES-128 
encryption (sample size = 256000 keys)}
\end{center}
\label{fig:aes-gaussian}
\vspace*{-0.2in}
\end{figure}

\begin{figure*}[t]
\begin{center}
\begin{tabular}{cccc}
\rotatebox{0}{
\includegraphics[scale = 0.35]{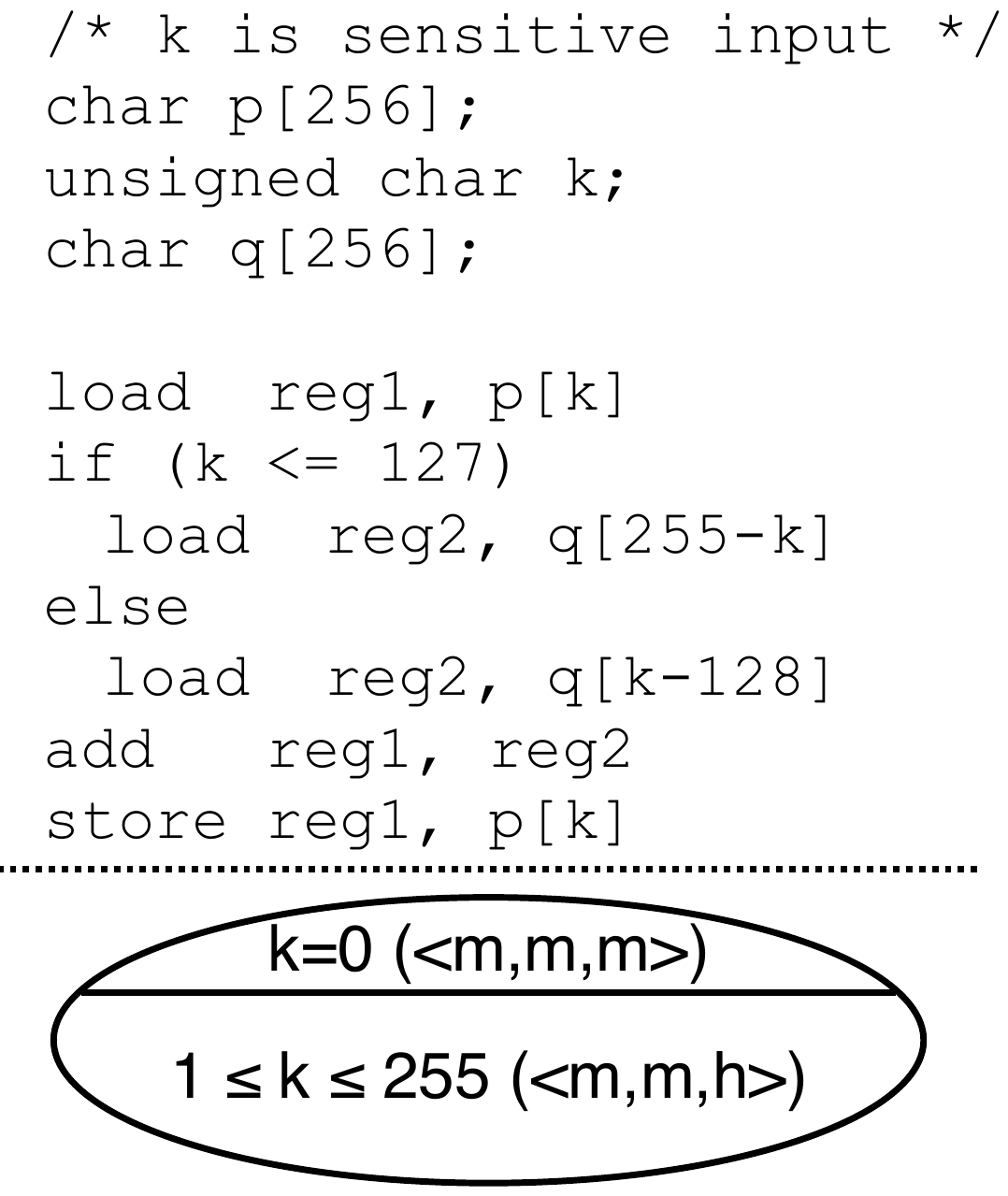}} &
\rotatebox{0}{
\includegraphics[scale = 0.35]{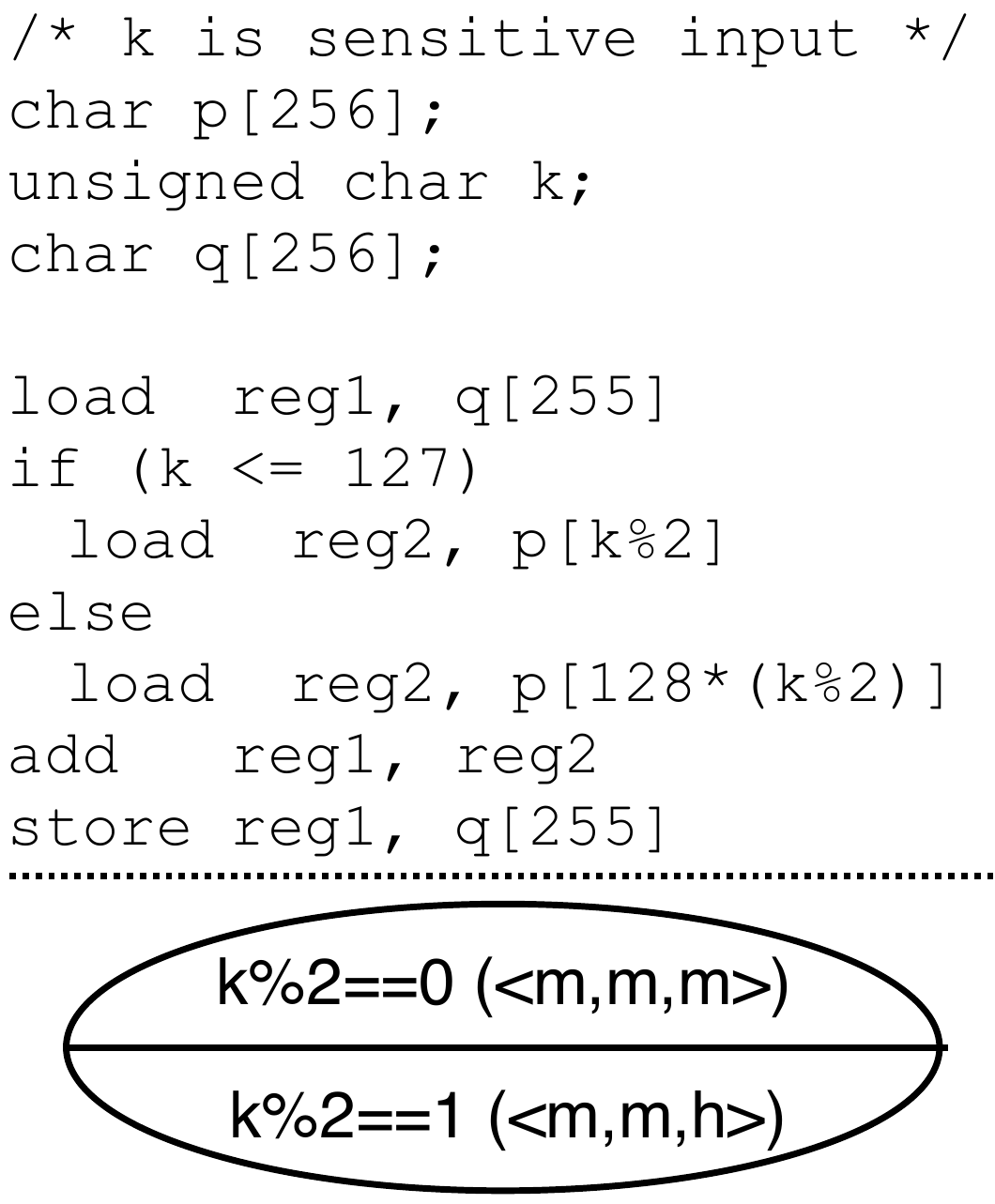}} & 
\rotatebox{0}{
\includegraphics[scale = 0.35]{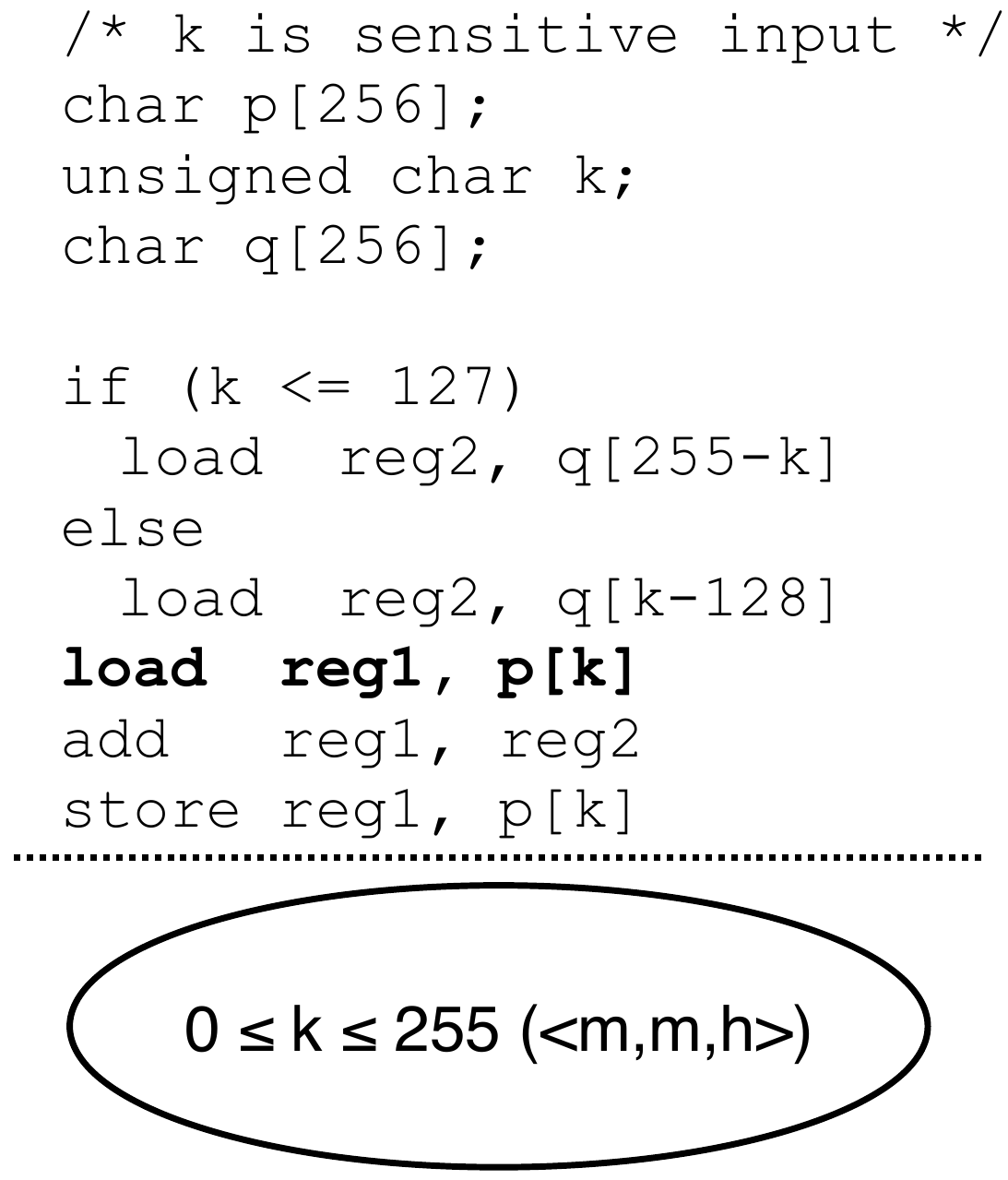}} & 
\rotatebox{0}{
\includegraphics[scale = 0.35]{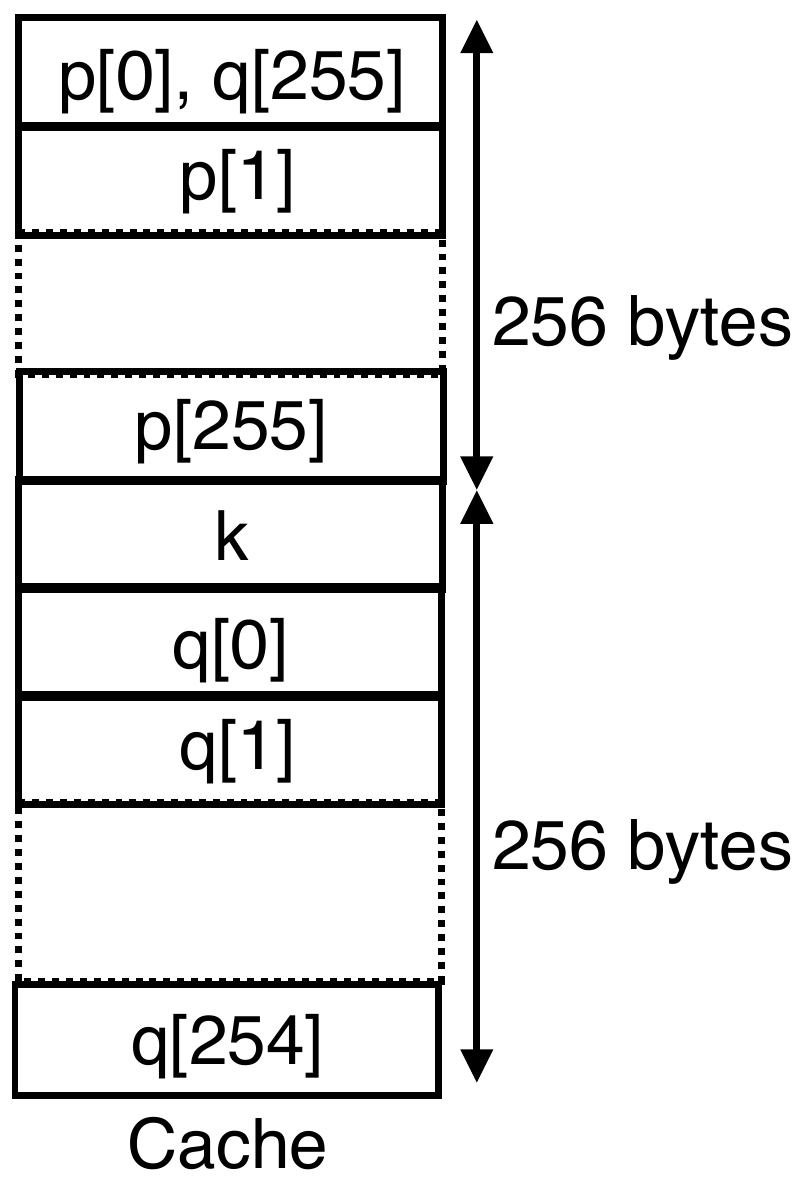}}\\
\textbf{(a)} & \textbf{(b)} & \textbf{(c)} & \textbf{(d)}
\end{tabular}
\end{center}
\vspace*{-0.1in}
\caption{$k$ is a sensitive input. (a)-(c) three code fragments and respective 
partitions of the input space with respect to cache hit/miss sequence ($reg1$, 
$reg2$ represent registers), (d) mapping of program variables into a direct-mapped 
cache sized 512 bytes ($q[255]$ and $p[0]$ conflict in the cache)}
\label{fig:moexample}
\end{figure*}

The reason why CHALICE works is that the timings of cache hits and misses are not uniformly distributed; and therefore, some specific timings may reveal more information than others.  Figure~\ref{fig:aes-gaussian} demonstrates the execution of an AES-128 implementation~\cite{aes-impl} for a fixed input and 256,000 different keys, inducing between 213~and~279 cache misses.  We see that the distribution of cache misses is essentially Gaussian; if the number of cache misses is average, there are up to 13,850 keys which induce this very cache timing.  If we have an extreme cache timing with 213~misses (the minimum) or 279~misses (the maximum), then there are only 2~keys that induce this very timing.  CHALICE can determine that for these keys, 90 of 128 bits would leak if the execution were subjected to a cache attack, which in practice would mean that the remaining 38 bits could be guessed through brute force---whereas other ``average'' keys would be much more robust. For each key and input, CHALICE \emph{can precisely predict which bits would leak,} allowing its users to determine and find the best alternative.\footnote{In the best of all worlds, one might have an implementation of every 
critical algorithm, such as an encryption routine, to have a uniform distribution over cache misses.  
But neither does such implementations exist that would be efficient, nor do we know whether such 
implementations \emph{can} exist; and replacing a well-studied algorithm like AES by some other 
algorithm with uniform distribution may induce other, yet unknown risks.}

It is this \emph{precision of its symbolic analysis} that sets CHALICE apart from the state of the art. 
Existing works~\cite{cacheaudit}~\cite{cav12-paper} use static analysis alone to provide an upper bound 
on the potential number of different observations that an attacker can make. This upper bound, however, 
does not suffice to choose between alternatives, as it ignores the \emph{distribution of inputs:}
It is possible that certain inputs may leak substantially more information than others. Not only that 
such an upper bound might be imprecise, it is also incapable to identify inputs that exhibits substantial 
information leakage through side channels. Given a set of inputs (typically as part of a testing pipeline), CHALICE can precisely quantify the leak for each input, and thus provide a full spectrum that  
characterizes inputs with respect to information leakage. 
%
%

The remainder of this paper is organized as follows.  After giving an overview on CHALICE (\autoref{sec:overview-example}), we make the following contributions:

\begin{enumerate}
\item We present CHALICE, \emph{a new approach to precisely quantify information leak in execution} and 
its usage in software testing (\autoref{sec:framework}).
\item We introduce \emph{a symbolic cache model} to instantiate CHALICE to detect 
cache side channel leakage (\autoref{sec:overall-method}).  This is the first usage of symbolic execution to explore cache states along 
with the program states.
\item We demonstrate that the model generalizes across \emph{multiple observer models.}  \autoref{sec:check-information-leak} demonstrates how CHALICE is instantiated
 for both direct-mapped and LRU cache replacement policies.
\item We provide an \emph{implementation} (\autoref{sec:implementation}) based on LLVM and the KLEE symbolic virtual machine.  Source code of CHALICE and all experimental data will be publicly available in the following URL: 
\begin{center}
\url{https://bitbucket.org/sudiptac/cache-side-channel}
\end{center}
\item We \emph{evaluate} our CHALICE approach (\autoref{sec:evaluation}) to show how we quantify the information leaked through 
execution in several libraries, including OpenSSL and Linux GDK libraries, and show that the information
leak can be as high as $251^{16}$ for certain implementations~\cite{aes-impl} of AES-128.
\end{enumerate}
After discussing related work (\autoref{sec:related-work}), we close with conclusion and consequences (\autoref{sec:conclusion}).
\newcommand{\program}{{\cal P}}
\newcommand{\proginput}{{\cal I}}
\newcommand{\progoutput}{{\cal O}}
\newcommand{\varinput}{{b_{in}}}
\newcommand{\varoutput}{{b_{out}}}

\newcommand{\predicates}{\Pi}
\newcommand{\predicate}{\pi}

\newcommand{\true}{\mathtt{true}}
\newcommand{\false}{\mathtt{false}}

\newcommand{\set}[1]{\left\{{#1}\right\}}

\newcommand{\observer}{\ensuremath{\mathcal{O}}}

\newcommand{\observe}{\ensuremath{\mathcal{C}}}

\newcommand{\leak}{\ensuremath{\mathcal{L}}}
\section{Overview}
\label{sec:overview-example}

In this section, we convey the key insight behind our approach through examples. 
In particular, we illustrate how \texttt{CHALICE} is used to quantify information 
leak from the execution trace of a program.

\subsection*{Motivating Example}
\label{sec:motivating-example}

Let us assume that our system contains a direct-mapped cache of size 512 bytes. 
Figures~\ref{fig:moexample}(a)-(c) show different code fragments executed 
in the system. 
For the sake of clarity, we use both assembly-level and source-level syntaxes. 
However, our framework takes a binary code as input, in order to accurately 
capture the memory behaviour of a program. For simplicity in the example, we 
assume that conditional checks 
do not involve any access to cache ({\em i.e.} $k$ is assigned to a register). 
%
The mapping of different variables into the cache is shown in 
Figure~\ref{fig:moexample}(d). Let us assume that the code fragments of 
Figures~\ref{fig:moexample}(a)-(c) are executed with some arbitrary (and unknown) 
value of $k$. Broadly, \texttt{CHALICE} answers the following 
question: {\em Provided only the cache performance (e.g. cache hit/miss sequence) 
from such executions, how much information about the sensitive input $k$ is leaked?} 


The cache performance induces a partition on the program input space. Let us capture 
the cache performance via a sequence of hits ($h$) and misses ($m$). 
In Figure~\ref{fig:moexample}(a), for all values of $k$ between 0 and 127, we observe 
two cache misses due to the first two memory accesses, $p[k]$ and $q[255-k]$, 
respectively. The second access to $p[k]$ is a {\em cache hit}, for $k \in [1,127]$. 
However, if $k=0$, the content of $p[k]$ will be replaced by $q[255-k]$, 
resulting in a cache miss at the second access of $p[k]$.
For $k \in [128,255]$, $p[k]$ is never replaced once it is loaded 
into the cache. Therefore, the second access to $p[k]$ is a cache hit for 
$k \in [128,255]$. In other words, we observe the sequence of cache hits and misses 
to induce the following partition on the input space: 
$k = 0$ (hit/miss sequence  = $\langle m,m,m \rangle$) and $k \in [1,255]$ (hit/miss 
sequence = $\langle m,m,h \rangle$). A similar exercise for the code in 
Figure~\ref{fig:moexample}(b) results in the following partition of the sensitive 
input space: $k \in [0,255] \wedge k\ mod\ 2 = 0$ (hit/miss sequence = 
$\langle m,m,m \rangle$) and $k \in [0,255] \wedge k\ mod\ 2 \ne 0$ (hit/miss sequence 
= $\langle m,m,h \rangle$). 

\subsubsection*{Key observation}
In this work, we stress the importance of {\em quantifying information leaks from 
execution traces and not from the static representation of a program}. To illustrate 
this, consider the input partitions created for code fragments in 
Figures~\ref{fig:moexample}(a)-(b). 
We emphasize that observing the cache hit/miss sequence $\langle m,m,m \rangle$, 
from an execution of the code fragment in Figure~\ref{fig:moexample}(a), results in 
complete disclosure of sensitive input $k$. On the contrary, observing the 
sequence $\langle m,m,m \rangle$, from an execution of the code fragment in 
Figure~\ref{fig:moexample}(b), will only reveal the information that {\em $k$ is odd}. 
Such information still demands a probability of $\frac{1}{128}$ in order to correctly 
guess $k$ at first attempt. This is in contrast to accurately guessing the 
correct value of $k$ at first attempt (as happened through the sequence 
$\langle m,m,m \rangle$ for Figure~\ref{fig:moexample}(a)). 
In order to fix the cache side-channel leak in Figure~\ref{fig:moexample}(a), we 
can reorder the code as shown in Figure~\ref{fig:moexample}(c).

\paragraph*{\textbf{Limitations of static analysis}}
Existing works in static analysis and verification have aimed at quantifying 
side channel leaks~\cite{cacheaudit,cav12-paper} and verifying 
constant-time implementations~\cite{ccs14-paper,usenix16-paper}. 
These works correlate the number of possible observations (by an attacker) 
with the number of bits leaked through a side channel. We believe this view 
can be dangerous.  
%
Indeed, both code fragments in Figures~\ref{fig:moexample}(a)-(b) have exactly 
two possible cache hit/miss sequences, for arbitrary values of $k$. 
Therefore, approaches based on static analysis~\cite{cacheaudit,cav12-paper} 
will consider these two code fragments {\em equivalent} in terms of cache 
side-channel leakage. As a result, statically analyzing a program will not reveal 
crucial information leak scenarios, such as the execution of code fragment in 
Figure~\ref{fig:moexample}(a) with $k=0$. 
Techniques based on verifying that programs execute in constant time typically 
check that memory accesses do not depend on sensitive inputs. Yet, most 
implementations do not execute in constant time. Besides, programs such as in 
Figure~\ref{fig:moexample}(c) have accesses that may depend on sensitive inputs 
without leaking information about it to a cache-performance observer. 
%
Therefore, we not only check the dependency between accessed memory address and 
program inputs, but we also accurately track the information flow through cache 
performance.


%

\subsubsection*{Can we use dynamic tainting?}
In the preceding paragraph, we state the importance of dynamically tracking sensitive 
information flow through cache performance. Approaches based on dynamic 
taints~\cite{taint-tracking-paper} can accomplish the task to detect information 
leak through standard functional outputs. However, such approaches fail to detect 
information leak through software non-functional outputs, such as cache performance, 
among others. Our methodology targets this angle of information leak detection by 
building a relationship between sensitive inputs and observed cache performance. 
In order to establish such a relationship, we leverage on symbolic analysis and 
constraint solving.

\subsubsection*{Limitations of side-channel vulnerability metrics}
In contrast to existing works on measuring cache side channel leakage~\cite{svf-paper}, 
we do not aim to check the strength of an attacker to {\em observe information through 
side channel}. Although promising, this work~\cite{svf-paper} {\em fails} to detect 
the information flow between sensitive inputs and observed performance. As a result, the 
side-channel vulnerability metric can only quantify {\em how well} an attacker can retrieve 
information from a system, but, {\em does not highlight the information potentially leaked 
to the attacker}. Of course, we believe our work is complementary to the metrics proposed 
in~\cite{svf-paper} and \texttt{CHALICE} could be combined with such metrics to build more 
advanced metrics for measuring side channel leakage. Such metrics could consider both 
information leaked by the system as well as the information that could be retrieved by 
an attacker.

\subsubsection*{The usage of \texttt{CHALICE}}
\texttt{CHALICE} is aimed to be used for validating security properties of software. 
Given a test suite ({\em i.e.} a set of concrete test inputs) for the software, 
\texttt{CHALICE} is used to quantify the information leaked for each possible 
observation obtained from this test suite. This is possible, as the observation by an 
attacker ({\em e.g.} number of cache miss) corresponds to a (set of) test inputs and 
\texttt{CHALICE} presents how much can be deduced about such inputs from the respective 
observation. 
In other words, our framework \texttt{CHALICE} fits the role of a 
{\em test oracle}~\cite{oracle} in the software validation process. For instance, 
if \texttt{CHALICE} reports substantial information leakage, the test inputs 
leading to the respective observation should be avoided ({\em e.g.} avoiding 
a ``weak" encryption key) or the candidate program needs to be restructured to 
avoid such information leak. 
The generation of an effective and optimized test suite, in order to detect cache side 
channels, is an open problem. However, \texttt{CHALICE} can be instantiated to generate 
a witness for each possible observations made by an attacker. The set of all these 
witnesses forms a concise test suite and our proposed method in \texttt{CHALICE} 
can quantify information leak for each element in such a test suite.
In this paper, we only focus on the quantification of information leak in a single 
test execution and not on the generation of a test suite. 

\texttt{CHALICE} {\em should not} be used for verifying the absence of cache 
side-channel leakage. Implementations that must adhere to zero-leakage, may 
leverage on \texttt{CHALICE} during the early design, specifically to discover 
the severity of potential cache side-channel leaks and the program locations 
exhibiting such leaks. Nevertheless, \texttt{CHALICE} is aimed for testing 
arbitrary software and we envision that such a strategy becomes an integral 
component of software testing pipeline in the future.

\subsubsection*{How \texttt{CHALICE} works}
%
%
Let us assume that we execute the code in Figure~\ref{fig:moexample}(a) with some 
input $I \in [0,255]$ and observed the trace $t_I \equiv \langle m,m,m \rangle$. 
{\em Given only the observation $t_I$, \texttt{CHALICE} quantifies how much 
information about program input $I$ is leaked.}
%
\texttt{CHALICE} symbolically executes the program and it tracks all memory accesses 
dependent on the sensitive input $k$. 
For each explored path, \texttt{CHALICE} constructs a symbolic cache model, which 
accurately encodes all possible cache hit/miss sequences for the respective path. 
In this example, \texttt{CHALICE} constructs $\Gamma (0 \leq k \leq 127)$ and 
$\Gamma (128 \leq k \leq 255)$, which encode all cache hit/miss sequences for 
inputs satisfying $0 \leq k \leq 127$ and $128 \leq k \leq 255$, respectively.
Let us consider the path explored for inputs $k \in [0,127]$. 
While exploring the path, we record a sequence of symbolic memory addresses 
$\langle \&p[k], \&q[255-k], \&p[k] \rangle$, where $\&x$ denotes the address 
of value $x$. Since we started execution with an empty cache, the first access to 
$p[k]$ inevitably incurs a cache miss, irrespective of the value of $k$. The subsequent 
accesses can either be cold misses (first access to the respective cache line) or eviction 
misses (non-first access to the respective cache line). Let us consider the second access 
to $p[k]$, as this is the memory access that partitions the input space. In order to check 
whether the second access to $p[k]$ is a cold miss, we check the following constraint: 
\begin{equation}
\label{eq:example1}
\begin{split}
\left ( 0 \leq k \leq 127 \right ) \wedge \left ( set(\&p[k]) \ne set(\&q[255-k]) \right )  
\\
\wedge
\left ( set(\&p[k]) \ne set(\&p[k]) \right )
\end{split}
\end{equation} 
where $set(\&x)$ captures the cache line where memory address $\&x$ is mapped to. Intuitively, 
the constraint checks whether access to $p[k]$ touches a cache line for the first time. 
Constraint~(\ref{eq:example1}) is clearly {\em unsatisfiable}, leading to the fact that 
the second access to $p[k]$ does not access a cache line for the first time during execution.

Subsequently, we check whether the second access to $p[k]$ can suffer an eviction miss. 
To this end, we check whether $q[255-k]$ can evict $p[k]$ from the cache as follows: 
\begin{equation}
\label{eq:example2}
\begin{split}
\left ( 0 \leq k \leq 127 \right ) \wedge \left ( set(\&p[k]) = set(\&q[255-k]) \right )  
\\
\wedge
\left ( tag(\&p[k]) \ne tag(\&q[255-k]) \right )
\end{split}
\end{equation} 
where $tag(\&x)$ captures the cache tag associated with the accessed memory block. Intuitively, 
Constraint~(\ref{eq:example2}) is satisfied if and only if $q[255-k]$ accesses a different 
memory block as compared to $p[k]$, but $q[255-k]$ and $p[k]$ access the same cache line 
(hence, causing an eviction before $p[k]$ was accessed for the second time). 
In this way, we collect Constraints~(\ref{eq:example1})-(\ref{eq:example2}) to formulate 
the cache behaviour of a memory access into $\Gamma (0 \leq k \leq 127)$.  

After constructing $\Gamma (0 \leq k \leq 127)$, we explore the path for inputs $k \in [128,255]$ 
and record the sequence of memory accesses $p[k]$, $q[k-128]$ and $p[k]$. Performing a similar 
exercise, we can show that the second access to $p[k]$ cannot be a cold miss along this path. 
In order to check whether the second access to $p[k]$ was an eviction miss along this path, we check 
whether $q[k-128]$ can evict $p[k]$ from the cache as follows:
\begin{equation}
\label{eq:example3}
\begin{split}
\left ( 128 \leq k \leq 255 \right ) \wedge \left ( set(\&p[k]) = set(\&q[k-128]) \right ) 
 \\
 \wedge
\left ( tag(\&p[k]) \ne tag(\&q[k-128]) \right )
\end{split}
\end{equation} 
Constraint~(\ref{eq:example3}) is used to formulate $\Gamma (128 \leq k \leq 255)$ 
and is unsatisfiable. This is because only $p[0]$ shares a cache line with $q[255]$ 
({\em i.e.} $set(\&p[0]) = set(\&q[255])$) and therefore, $set(\&p[k]) = set(\&q[k-128])$ 
is evaluated {\em false} for $128 \leq k \leq 255$. 
As a result, the second access to $p[k]$ is not a cache miss for any input 
$k \in [128,255]$. 

From the observation $\langle m,m,m \rangle$, we know that the second access to $p[k]$ was 
a miss. From the discussion in the preceding paragraph, we also know that this observation 
cannot occur for any inputs $k \in [128,255]$. 
Therefore, the value of $k$ must result in Constraint~(\ref{eq:example2}) satisfiable. 
Constraint~(\ref{eq:example2}) is unsatisfiable if we restrict the value of $k$ between 
1 and 127. This happens based on the fact that only $p[0]$ is mapped to the same cache 
line as $q[255]$ ({\em cf.} Figure~\ref{fig:moexample}(d)). As a result, \texttt{CHALICE} 
reports 255 (127 for the \texttt{if} branch and 128 for the \texttt{else} branch in 
Figure~\ref{fig:moexample}(a)) values being leaked for the observation $\langle m,m,m \rangle$.  
In other words, \texttt{CHALICE} accurately reports the information leak ({\em i.e.} $k=0$) 
for the observation $\langle m,m,m \rangle$.



\section{Framework}
\label{sec:framework}

In the following, we formally introduce the problem statement and provide 
an outline of our overall approach to solve this problem. 

\subsection{Foundation}

\subsubsection*{Threat model}
Side-channel attacks are broadly classified into synchronous and asynchronous 
attacks~\cite{tromer-paper}. In synchronous attack, an attacker can trigger 
the processing of known inputs ({\em e.g.} a plain-text or a cipher-text for 
encryption routines), whereas such a phenomenon is not possible for asynchronous 
attacks. Synchronous attacks are clearly easier to perform, since the attacker 
does not need to compute the start and end of the targeted routine under 
attack. For instance, in synchronous attack, the attacker can trigger encryption 
of known plaintext messages and observe the encryption-timing~\cite{djb:2005cache}. 
Since \texttt{CHALICE} is a software validation tool with the aim of producing 
side-channel resistant implementations, we assume the presence of a strong 
attacker in this paper. Therefore, we consider the attacker can request and 
observe the execution ({\em e.g.} number of cache miss) of the targeted routine. 
We also assume that the attacker can execute arbitrary user-level code in the 
same processor running the targeted routine. This allows the attacker to flush 
the cache before the targeted routine starts execution and therefore, reduce the 
external noise in the observation. The attacker, however, is incapable to 
access the address space of the target routine.  

\subsubsection*{Notations}
%
The execution of program $\program$ on input $I$ results in an execution 
trace $t_I$. $t_I$ is a sequence over the alphabet $\Sigma=\{h,m\}$ where 
$h$ (respectively, $m$) represents a cache hit (respectively, cache miss). 
%
Our proposed method in \texttt{CHALICE} quantifies the information leaked 
through $t_I$. We capture this quantification via $\leak(t_I)$. We assess 
the information leakage with respect to an {\em observer}.
An {\em observer} is a mapping $\observer:\Sigma^*\to \mathbb{D}$ where 
$\mathbb{D}$ is a countable set. For instance, an observer 
$\observer:\Sigma^*\to\mathbb{N}$ can count the number of misses and will 
associate both sequences $\langle m, h, m, h, h \rangle$ and 
$\langle m, m, h, h, h \rangle$ to $2$. It will therefore not differentiate 
them. The most precise observer would be the identity mapping on
$\Sigma^*$. However, an observer that tracks prefixes of some fixed lengths 
(for example 2) would be enough to differentiate the two aforementioned 
sequences. 

We use the variable $miss_i$ to capture whether or not the $i$-th memory 
access was a cache miss during execution. The observation by an attacker, 
over the execution for an arbitrary input and according to the observer model 
$\observer$, is considered via the observation constraint $\Phi_{\observer}$. 
$\Phi_{\observer}$ is a symbolic constraint over the set of variables 
$\{miss_1,miss_2,\ldots,miss_n\}$, where $n$ is the total number of memory 
accesses during an execution. For instance, 
$\Phi_{\observer} \equiv \left ( \sum_{i=1}^{n}\ miss_i = 100 \right )$ 
accurately captures that the attacker observes 100 cache misses in an execution 
manifesting $n$ memory accesses. 
For the sake of formulation, we use $\Phi_{\observer,e}$ to define a 
{\em projection} of $\Phi_{\observer}$ on an arbitrary program path 
$e$. In particular, $\Phi_{\observer,e}$ captures the observation constraint 
if program path $e$ is executed. 
Given only $\Phi_{\observer}$ to be observed by an attacker, \texttt{CHALICE} 
quantifies how much information about the respective program input is leaked.  

The central idea of our information leak detection is to capture the 
cache behaviour via symbolic constraints. Let us consider a set of 
inputs $\mathbb{I}$ that exercise the same execution path with $n$ memory 
accesses. We use $\Gamma(\mathbb{I})$ to accurately encode all possible 
combinations of values of variables $\{miss_1,miss_2,\ldots,miss_n\}$. 
Therefore, if $\Gamma(\mathbb{I}) \wedge \Phi_{\observer}$ is {\em unsatisfiable}, 
we can deduce that the respective observation $\Phi_{\observer}$ 
{\em did not occur} for any input $I \in \mathbb{I}$.   

We now describe how $\leak(t_I)$ is computed based on the notations and the 
intuition mentioned in the preceding.


\begin{figure}[t]
\begin{center}
\rotatebox{0}{
\includegraphics[scale = 0.27]{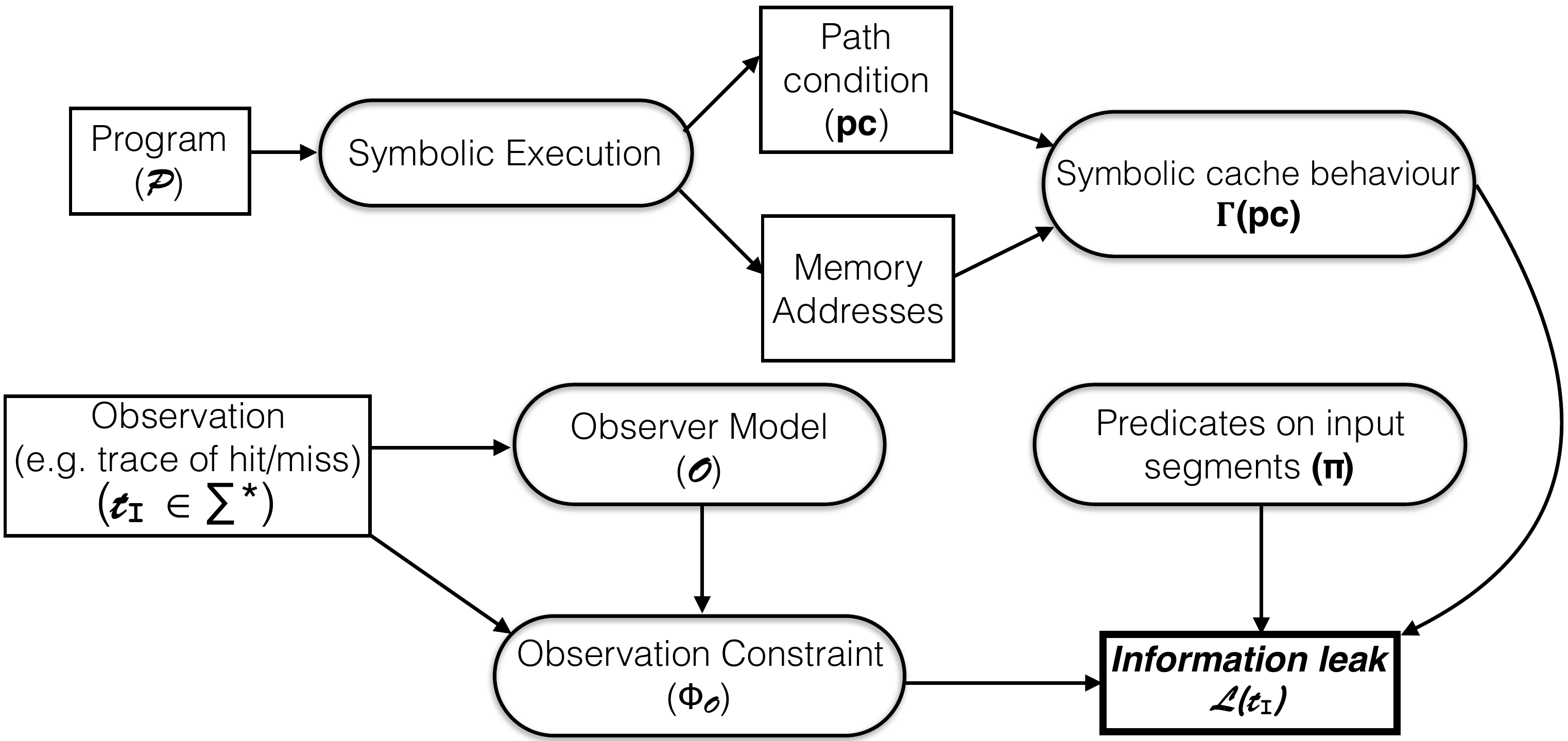}}
\vspace*{-0.1in}
\caption{The framework \texttt{CHALICE}.}
\end{center}
\label{fig:overview}
\vspace*{-0.2in}
\end{figure}

\subsection{Quantifying Information Leak in Execution}
\label{sec:approach}

Figure~\ref{fig:overview} provides an outline of our entire framework.
We symbolically execute a program $\program$ compute the path 
condition~\cite{dart-paper} for each explored path. Such a path condition 
symbolically encodes all program inputs for which the respective program 
path was followed.  
Our symbolic execution based framework tracks all memory accesses on
a taken path and therefore, enables us to characterize, for all symbolic 
arguments satisfying the path condition, the set of all associated cache 
behaviors.

Recall that we use $\Gamma (\mathbb{I})$ to capture possible cache 
hit/miss sequences in an execution path, which was activated by a set 
of inputs $\mathbb{I}$. In an abuse of notation, we capture set of 
inputs $\mathbb{I}$ via path conditions. For instance, 
in Figure~\ref{fig:moexample}(a), we use $\Gamma (0 \leq k \leq 127)$ 
to encode all possible cache hit/miss sequences for inputs activating 
the \texttt{If} branch. 

For an arbitrary execution path, let us consider $pc$ be the path condition. 
Along this path, we record each memory access and we consider its cache 
behaviour via variable $miss_i$. $miss_i$ is set to 1 (resp. 0) if and only 
if the $i$-th memory access along the path encounters a cache miss (hit). 
Given $n$ to be the total number of memory accesses along the path, we 
formulate $\Gamma (pc)$ to bound the value of $\{miss_1, miss_2, \ldots, miss_n\}$. 
In particular, any solution of $\Gamma (pc) \wedge \left ( miss_i = 1 \right )$ 
captures a concrete input $I \Rightarrow pc$ and such an 
input $I$ leads to an execution where the $i$-th memory access is a cache miss. 
%
%
%
Therefore, if an observation $\Phi_{\observer}$ happens to be for input 
$I \Rightarrow pc$, $\Gamma (pc) \wedge \Phi_{\observer}$ is always satisfiable.

We capture the information leak through execution trace $t_I$ as follows: 
\vspace*{-0.07in}
\begin{equation}
\label{eq:generic-leak}
\boxed{
\leak(t_I) = 2^{N} - 
|\bigvee_{e \in \mathit{Path}} \left ( \Gamma (pc_e) \wedge \Phi_{\observer , e} \right ) |_{sol}}
\end{equation}
where $N$ is size of program input (in bits), $\Phi_{\observer , e}$ is the 
projection of the observation constraint on path $e$, $\mathit{Path}$ is the 
set of all program paths and $pc_e$ is the path condition for program path $e$. 
$|\mathcal{X}|_{sol}$ captures the number of solutions satisfied by 
predicate $\mathcal{X}$. It is worthwhile to note that 
$|\bigvee_{e \in \mathit{Path}} \left ( \Gamma (pc_e) \wedge \Phi_{\observer , e} \right ) |_{sol}$ 
accurately captures the number of program inputs that exhibit the observation 
satisfied by $\Phi_{\observer}$. In other words, Equation~(\ref{eq:generic-leak}) 
quantifies the number of program inputs that does not exhibit the observation, 
as captured by $\Phi_{\observer}$. Hence, if the attacker 
observes $\Phi_{\observer}$, she can deduce as many as $\leak(t_I)$ inputs were 
impossible for the respective observation.

In practice, however, computing the exact value of $\leak(t_I)$ 
might be infeasible, as it might require the enumeration of all solutions. 
In order to control such enumeration, we generate predicates on input 
variables. In particular, we sample an $N$-bit input into $K$ equal 
segments, resulting in input segments of length $\frac{N}{K}$. 
Subsequently, we constrain the search space of the solver by 
restricting the value of each such input segment to any possible 
value, that is, pointing to a value in the set 
$\{0,1,\ldots,2^{\frac{N}{K}} - 1\}$. 
For instance, let us assume $x$ is the program input and $x_i$ captures the 
$i$-th input segment. A predicate 
$\pi \equiv \left ( x_i = 0 \right )$ will guide the solver to search 
for a solution only in the input space where the $i$-th input segment 
is 0. Since, we have $K$ different segments, we generate a total of 
$\left ( K \cdot 2^{\frac{N}{K}} \right )$ different predicates. 
For each such predicate $\pi$, we record information leak if the 
following constraint is {\em unsatisfiable}: 
%
\begin{equation}
\label{eq:heuristic-leak}
\bigvee_{e \in \mathit{Path}} \left ( \Gamma (pc_e) \wedge \Phi_{\observer , e} \wedge \pi \right )
\end{equation}
Concretely, if Constraint~(\ref{eq:heuristic-leak}) is unsatisfiable, we can 
accurately record that input $I$, which leads to observation $\Phi_{\observer}$ 
along some program path, satisfies the predicate $\neg \pi$. 
An appealing feature of this process is that all $K \cdot 2^{\frac{N}{K}}$ 
predicates can be generated independently and therefore, the unsatisfiability 
check of Constraint~(\ref{eq:heuristic-leak}) can be performed in parallel 
for different predicates.  

Let us assume, $\mathcal{U}_1, \mathcal{U}_2, \ldots, \mathcal{U}_K$ are the 
number of unsatisfiable solutions reported for each of the $K$ input segments 
respectively. Therefore, we can estimate a lower bound on $\leak(t_I)$ from 
these unsatisfiability checks as follows: 
\begin{equation}
\label{eq:heuristic-leak-bound}
\boxed{
\leak(t_I) \geq 2^N - \prod_{1 \leq i \leq K} \left ( 2^{\frac{N}{K}} - \mathcal{U}_i \right )}
\end{equation}

Due to the classic path explosion problem in symbolic execution, it is possible 
that only a subset of paths $\mathcal{P}' \subseteq \mathit{Path}$ can be explored 
within a given time budget. In such cases, we can quantify $\leak(t_I)$ as follows. 
\begin{equation}
\label{eq:heuristic-leak-bound-partial}
\begin{split}
\leak(t_I) & = 2^{N} - 
|\bigvee_{e \in \mathit{Path}} \left ( \Gamma (pc_e) \wedge \Phi_{\observer , e} \right ) |_{sol}
\\
& = |\bigvee_{e \in \mathcal{P}'} pc_e|_{sol} + 
|\bigvee_{e \in \mathit{Path} \setminus \mathcal{P}'} pc_e|_{sol}
\\
& - |\bigvee_{e \in \mathcal{P}'} \left ( \Gamma (pc_e) \wedge \Phi_{\observer , e} \right ) |_{sol} 
\\
& - |\bigvee_{e \in \mathit{Path} \setminus \mathcal{P}'} 
\left ( \Gamma (pc_e) \wedge \Phi_{\observer , e} \right ) |_{sol}
\\
& \geq |\bigvee_{e \in \mathcal{P}'} pc_e|_{sol} 
- |\bigvee_{e \in \mathcal{P}'} \left ( \Gamma (pc_e) \wedge \Phi_{\observer , e} \right ) |_{sol}
\\
& \boxed {\geq |\bigvee_{e \in \mathcal{P}'} pc_e|_{sol} - \prod_{1 \leq i \leq K} \left ( 2^{\frac{N}{K}} - \mathcal{U}_i \right )}
\end{split}
\end{equation}
This result follows from the fact that 
$\Gamma (pc_e) \wedge \Phi_{\observer , e} \Rightarrow \Gamma(pc_e) \Rightarrow pc_e$. 
The term $|\bigvee_{e \in \mathcal{P}'} pc_e|_{sol}$ involves only path 
conditions and it can be computed via model counting~\cite{model-counting}. 

Finally, it is worthwhile to note that setting $\mathit{K=1}$ is equivalent to 
enumerating all solutions as in Equation~(\ref{eq:generic-leak}). In contrast, 
setting $\mathit{K=N}$ is equivalent to checking information leak at bit-level 
({\em i.e.} checking whether the value of a single bit can influence cache 
performance). Therefore, $K$ provides a tunable parameter for different levels 
of information leak detection. We have conducted evaluation for $\mathit{K=8}$ 
and $\mathit{K=N}$. This means, we have checked how much information about a 
single byte and respectively, a single bit are leaked through observing cache 
performance.

In the next section, we will describe the construction of $\Gamma \left ( pc \right )$ 
for an arbitrary path condition $pc$.

\section{Generating Symbolic Cache Model}
\label{sec:overall-method}
The technical contribution of our methodology is a symbolic model for 
cache behaviour -- establishing a link between the program input and 
observed cache performance. 
To describe our model, we shall use the following notations throughout 
our discussions:
\begin{itemize}
 \item $2^{\mathcal{S}}:$ The number of cache sets in the cache. 
 \item $2^{\mathcal{B}}:$ The size of a cache line (in bytes). 
 \item $\mathcal{A}:$ Associativity of cache. For direct-mapped caches, $\mathcal{A}=1$.	
 \item $set(r_i):$ Cache set accessed by instruction $r_i$. 
 \item $tag(r_i):$ The tag stored in the cache for the memory block accessed by $r_i$. 
 \item $\zeta_i:$ The cache state before executing instruction $r_i$ and after executing
 instruction $r_{i-1}$. 
\end{itemize}
In the following, we will explain the different steps of generating 
the symbolic cache model.

\subsection{Intercepting Memory Requests}
\label{sec:intercept-mem-request}

We symbolically execute a program $P$. During symbolic execution, we 
track the path condition and the sequence of memory accesses for each 
explored path. For instance, while symbolically exercising the 
\texttt{If} branch of Figure~\ref{fig:moexample}(a), we track the path 
condition $0 \leq k \leq 127$ and the sequence of memory addresses 
$\langle \&p[k], \&q[255-k], \&p[k] \rangle$. It is worthwhile to note 
that such memory addresses might capture symbolic expressions due to 
the dependency from program inputs. Concretely, we compute the path 
condition $pc$ and the execution trace $\Psi_{pc}$ for each explored 
path as follows: 
\begin{equation}
\label{eq:memory-address}
\Psi_{pc} \equiv \langle (r_1, \sigma_1), (r_2, \sigma_2), \ldots, 
(r_{n-1}, \sigma_{n-1}), (r_n,\sigma_n) \rangle
\end{equation}
where $r_i$ captures the $i$-th memory-related instruction executed 
along the path and $\sigma_i$ symbolically captures the memory address 
accessed by $r_i$. 


\subsection{Modeling Symbolic Cache Access}
\label{sec:symbolic-cache-access}
In order to find the impact on caches, we need to find out the set 
of cache lines being accessed. This is accomplished by manipulating the 
expression $\sigma_i$, which was collected while executing each 
memory-related instruction $r_i$ ({\em cf.} Equation~(\ref{eq:memory-address})).  
In particular, we formulate $set(r_i)$ as follows:
\begin{equation}
\label{eq:set-expression}
set(r_i) = \left ( \sigma_i \gg \mathcal{B} \right )\ \& \ \left ( 2^\mathcal{S} - 1 \right )
\end{equation}
In Equation~(\ref{eq:set-expression}), ``\texttt{\&}'' captures a bitwise-and operation 
and ``$\gg$" captures a right-shift operation. 

Apart from the cache set a memory address is mapped to, we need to distinguish different 
memory addresses from which contents are stored into the cache. This is different from 
just checking the inequality between $\sigma_i$ values, as the memory controller groups 
contents of different memory addresses into a {\em memory block} and stores the memory 
block into a cache line. 
In order to distinguish different memory blocks mapped into the same cache lines, a 
tag is stored within each cache line. For instruction $r_i$, such a tag $tag(r_i)$ 
is captured as follows:
\begin{equation}
\label{eq:tag-expression}
tag(r_i) = \left ( \sigma_i \gg \left ( \mathcal{B} + \mathcal{S} \right ) \right )
\end{equation}
Therefore, if $tag(r_i) \ne tag(r_j)$, we can conclude that $r_i$ and $r_j$ are accessing 
different memory blocks, even if $set(r_i) = set(r_j)$ holds. 

It is worthwhile to note that both $set(r_i)$ and $tag(r_i)$ might be symbolic expressions 
due to the presence of symbolic expression $\sigma_i$ 
in Equations~(\ref{eq:set-expression})-(\ref{eq:tag-expression}). 
Moreover, the computations of $set(r_i)$ and $tag(r_i)$ are independent of any 
cache replacement policy.

\subsection{Direct-mapped Caches}
\label{sec:direct-mapped}
In this section, we assume that the cache is {\em direct-mapped}. Therefore, 
each cache set holds exactly one cache line. In the next section, we extend 
our symbolic model for set-associative caches. 

We characterize cache misses into the following two categories:
\begin{enumerate}
  \item Cold cache misses. $r_i$ suffers a cold miss 
 {\em if and only if} $set(r_i)$ has not been accessed by any previous 
 instruction $r \in \{r_1, r_2, \ldots, r_{i-1}\}$.  	

 \item Cache misses due to eviction. $r_i$ suffers a cache miss due to 
 eviction
 {\em if and only if} the last access to $set(r_i)$ had been from an 
 instruction $r_j \in \{r_1, r_2, \ldots, r_{i-1}\}$, such that 
 $tag(r_j) \neq tag(r_i)$.
  
\end{enumerate}


\begin{figure}
\begin{center}
\includegraphics[scale = 0.33]{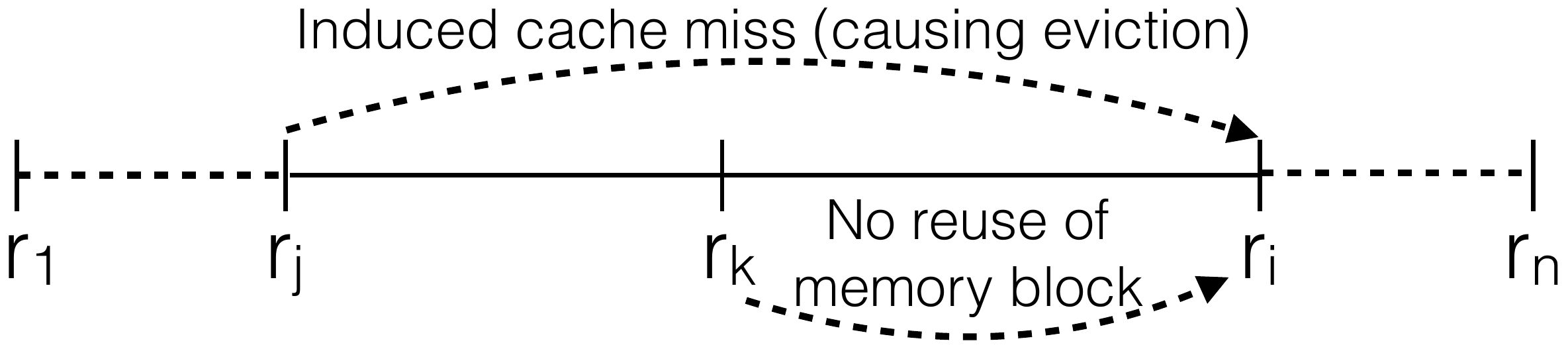}
\end{center}
\vspace*{-0.2in} 
\caption{Memory-access $r_j$ induces a cache miss at $r_i$ if $r_j$ accesses 
the same cache set as $r_i$ and $r_k$ does not load the block accessed 
by $r_i$}
\label{fig:example}
\vspace*{-0.1in} 
\end{figure}


\subsubsection*{Constraints to formulate cold cache misses}
If a cache line is accessed for the {\em first time}, such an access will 
inevitably incur a cache miss. 
%
%
Let us consider that we want to check whether instruction $r_i$ accesses 
a cache line for the first time during execution. 
In other words, we can check none of the instruction 
$r \in \{r_1, r_2, \ldots, r_{i-1}\}$ touches the same cache line as $r_i$. 
Therefore $r_i$ suffers a cold miss if and only if the following condition 
holds:
%
\begin{equation}
\label{eq:cold-miss-sequence-pre}
\Theta_i^{cold} \equiv \bigwedge_{p \in [1,i)} \left ( set(r_p) \ne set(r_i) \right )
\end{equation}



\subsubsection*{Constraints to formulate cache evictions}
\label{sec:evict-formulation}
In the following, we formulate a set of constraints to encode cache 
misses other than cold cache misses. Such cache misses occur due 
to the eviction of memory blocks from caches.

To illustrate different cache-miss scenarios clearly, let us consider 
the example shown in Figure~\ref{fig:example}. Assume that we want to 
check whether $r_i$ will suffer a cache miss due to eviction. This 
might happen only due to instructions appearing before 
(in the program order) $r_i$. Consider one such instruction $r_j$, for 
some $j \in [1,i)$. Informally, $r_j$ is responsible for a cache 
miss at $r_i$,  {\em only if} the following conditions hold:
\begin{enumerate}
 \item $\psi_{cnf}(j,i)$: $r_i$ and $r_j$ access the same cache set. 
 Therefore, we have the following constraint: 
 \begin{equation}
 \label{eq:set-constraint}
 \psi_{cnf} \left ( j,i \right ) \equiv \left ( set(r_j) = set(r_i) \right )
 \end{equation}

 \item $\psi_{dif}(j,i)$: $r_i$ and $r_j$ access different memory-block 
 tags. This can be formalized as follows:
 \begin{equation}
 \label{eq:tag-constraint}
 \psi_{dif} \left ( j,i \right ) \equiv \left ( tag(r_j) \ne tag(r_i) \right )
 \end{equation}

 \item $\psi_{eqv}(j,i)$: There does not exist any instruction $r_k$ 
 where $k \in [j+1,i)$, such that $r_k$ accesses the same memory block 
 as $r_i$. It is worthwhile to note that the existence of $r_k$ will 
 load the memory block accessed at $r_i$. Since $r_k$ is executed after 
 $r_j$ (in program order), $r_j$ must not be responsible for a cache 
 miss at $r_i$. We formulate the following constraint to capture this 
 condition: 
 \begin{equation}
\label{eq:equiv-constraint}
\begin{split}
\psi_{eqv} \left ( j,i \right ) \equiv   
\bigwedge_{k:\ j < k < i} \left ( tag(r_k) \ne tag(r_i) \right.
\\
\vee \left. set(r_k) \ne set(r_i) \right ) 
\end{split}
\end{equation}
\end{enumerate}

Constraints~(\ref{eq:set-constraint})-(\ref{eq:equiv-constraint}) capture necessary 
and sufficient conditions for instruction $r_j$ to replace the memory block 
accessed by $r_i$ (where $j < i$) and the respective block not being accessed between 
$r_j$ and $r_i$. In order to check whether $r_i$ suffers a cache miss due to eviction, 
we need to check Constraints~(\ref{eq:set-constraint})-(\ref{eq:equiv-constraint}) 
for any $r \in \{r_1,r_2,\ldots,r_{i-1}\}$. This can be captured via the 
following constraint:
\begin{equation}
\label{eq:evict-one-miss-pre}
\Theta_i^{emp} \equiv \left ( \bigvee_{j:\ 1 \leq j < i} \left (\psi_{cnf} \left ( j,i \right ) 
\wedge 
\psi_{dif} \left ( j,i \right ) \wedge \psi_{eqv} \left ( j,i \right ) \right ) \right )
\end{equation}

$r_i$ will not suffer a cache miss due to eviction when at least one of the Constraints~(\ref{eq:set-constraint})-(\ref{eq:equiv-constraint}) 
does not hold for all prior instructions of $r_i$. 
This scenario is the negation of Constraint~(\ref{eq:evict-one-miss-pre}) and therefore, 
it is captured via $\neg \Theta_i^{emp}$.
%

We use variable $miss_i$ to capture whether instruction $r_i$ suffers a cache miss. 
As discussed in the preceding paragraphs, $r_i$ suffers a cold miss 
({\em i.e.} satisfying Constraint~(\ref{eq:cold-miss-sequence-pre})) 
or the memory block accessed by $r_i$ would be evicted due to instructions executed 
before $r_i$ ({\em i.e.} satisfying Constraint~(\ref{eq:evict-one-miss-pre})). 
Using this notion, we formulate the value of $miss_i$ as follows:
\begin{equation}
\label{eq:evict-one-miss-precondition}
\Theta_i^{mp,dir} \equiv \left ( \Theta_i^{emp} \vee \Theta_i^{cold} \right )
\end{equation}
\begin{equation}
\label{eq:evict-one-miss}
\Theta_i^{m,dir} \equiv \left ( \Theta_i^{mp,dir} \Rightarrow \left ( miss_i = 1 \right ) \right )
\end{equation}
\begin{equation}
\label{eq:evict-one-hit}
\Theta_i^{h,dir} \equiv \left ( \neg \Theta_i^{mp,dir} \Rightarrow \left ( miss_i = 0 \right ) \right )
\end{equation}



\paragraph*{\textbf{Putting it all together}}
Recall that $\Gamma(pc)$ captures the constraint system to encode the 
cache behaviour for all inputs $I \Rightarrow pc$. In order to construct 
$\Gamma(pc)$, we gather constraints, as derived in the preceding 
sections, and the path condition into $\Gamma(pc)$ as follows:
\begin{equation}
\boxed{
\label{eq:gamma-direct-miss-count}
\Gamma(pc) \equiv pc \wedge \bigwedge_{i \in [1,n]} \left 
( \Theta_i^{m,dir} \wedge \Theta_i^{h,dir} \right ) }
\end{equation}

\subsection{Set-associative Caches}
\label{sec:lru-cache}

In direct-mapped caches, exactly one memory-block tag is contained by a cache 
set. As a result, this memory block is replaced by any instruction accessing 
the same cache set, but accessing a different memory-block tag. In contrast, 
set-associative caches group multiple cache lines into a cache set. 
Therefore, evicting a memory block from a cache set might require multiple 
accesses to the respective cache set. 
The number of such accesses, as required to evict a memory block from a cache 
set, is determined by the relative position of the same block within the cache 
set. This relative position is updated during execution according to a cache 
replacement policy. In this paper, we instantiate \texttt{CHALICE} for 
set-associative caches with LRU replacement policy.

From technical perspective, we need to modify 
Constraints~(\ref{eq:evict-one-miss-pre})-(\ref{eq:evict-one-hit}) to reflect 
the working principle of set-associative caches. 
Before discussing such modifications, we introduce the concept of {\em cache 
conflict}, which is crucial for formulating the cache behaviour of 
set-associative caches.  

\begin{definition}
\textbf{(Cache Conflict):} $r_j$ generates a cache conflict to $r_i$
only if executing $r_j$ can influence the relative position of memory
block accessed by $r_i$ within the cache state $\zeta_i$ ({\em i.e.} 
the cache state before $r_i$ and after $r_{i-1}$).
\end{definition}


In order to check whether $r_i$ suffers a cache miss, we distinguish between 
the following two scenarios:
\begin{enumerate}
\item $r_i$ accesses a memory block for the first time. Hence, $r_i$ will suffer 
a cold cache miss. 
\item The number of unique cache conflicts generated to $r_i$ is sufficient to 
evict the memory block accessed by $r_i$. Hence $r_i$ will suffer a cache miss.  
\end{enumerate}

\subsubsection*{Constraints to formulate cold cache misses}
If $r_i$ accesses a memory block for the first time, the following condition must hold:
\begin{equation}
\label{eq:cold-miss-set}
\begin{split}
\Theta_{i}^{cold} \equiv \bigwedge_{1 \leq k < i} 
\left ( tag \left ( r_k \right ) \ne tag \left ( r_i \right ) \right.
\\
\vee 
\left. set \left ( r_k \right ) \ne set \left ( r_i \right ) \right )
\end{split}
\end{equation}
Informally, Constraint~(\ref{eq:cold-miss-set}) states that every instruction 
$r \in \{r_1,r_2, \ldots, r_{i-1}\}$ either accesses a different cache set 
than $set(r_i)$ or the accessed memory block has a different tag compared to 
$tag(r_i)$. This leads to a cold cache miss at $r_i$.

\subsubsection*{Constraints to formulate cache evictions}
The eviction of a memory block from the cache is critically influenced by cache 
conflict. Therefore, we need to consider all scenarios where a cache conflict 
might be generated.   
For LRU caches, $r_j$ generates a cache conflict to $r_i$ (where $j < i$) only if 
the following conditions hold: 
\begin{enumerate}
\item $\psi_{cnf}(j,i)$, $\psi_{dif}(j,i)$ and $\psi_{eqv}(j,i)$ hold ({\em cf.} 
Constraints~(\ref{eq:set-constraint})-(\ref{eq:equiv-constraint})). This ensures 
that $r_j$ and $r_i$ access the same cache set, but different memory-block tags. 
Additionally, $\psi_{eqv}(j,i)$ ensures that there does not exist any instruction 
between $r_j$ and $r_i$ that loads the memory block accessed by $r_i$. 

\item Note that multiple accesses may influence the cache content in set-associative 
caches. Therefore, we need to distinguish unique memory accesses in order to formulate 
cache conflict. 
For instance, consider the following 
memory accesses in sequence: $r_1$:$m_1 \rightarrow r_2$:$m_2 \rightarrow r_3$:$m_2 
\rightarrow r_4$:$m_1$, where $r_i$ captures the instruction and $m_j$ captures 
the respective memory block being accessed. If $m_1$ and 
$m_2$ map to the same cache set in a 2-way LRU cache, $r_4$ will still be a 
cache hit. This is because $r_4$ suffers cache conflict only once, from the 
access to memory block $m_2$, even though $m_2$ has been accessed twice 
(at $r_2$ and at $r_3$). In order to account unique cache conflicts, we only 
record the cache conflict from the {\em closest} access to different memory 
blocks. For instance, in the preceding example, we only record cache 
conflict from $r_3$ to $r_4$. Formally, we need additional constraints to 
distinguish such closest accesses. We use the constraint 
$\psi_{unq} \left ( j,i \right )$ for such purpose. 
$\psi_{unq} \left ( j,i \right )$ is satisfiable if and only if there does not 
exist any instruction between $r_j$ (where $j \in [1,i)$) and $r_i$ that 
accesses the same memory block as $r_j$. $\psi_{unq} \left ( j,i \right )$ is 
formalized as follows:
\begin{equation}
\label{eq:equiv-set-constraint}
\begin{split}
\psi_{unq} \left ( j,i \right ) \equiv 
\bigwedge_{k:\ j < k < i} \left ( tag(r_j) \ne tag(r_k) \right. 
\\
\vee \left. set(r_j) \ne set(r_k) \right ) 
\end{split}
\end{equation}
\end{enumerate}

Constraints~(\ref{eq:set-constraint})-(\ref{eq:equiv-constraint}) 
and Constraint~(\ref{eq:equiv-set-constraint}) accurately capture scenarios where 
$r_j$ ($j \in [1,i)$) will create a unique cache conflict to $r_i$. Let us assume 
$\Psi_{i,j}^{evt}$ captures whether $r_j$ creates a unique cache conflict to $r_i$. 
Using the intuition described in the preceding paragraph, we can formulate the 
following constraints to set the value of $\Psi_{i,j}^{evt}$.
\begin{align}
\label{eq:evict-one-miss-set}
\Theta_{j,i}^{em} \equiv \left ( \psi_{cnf} \left ( j,i \right ) 
\wedge 
\psi_{dif} \left ( j,i \right ) \wedge \psi_{eqv} \left ( j,i \right ) \right.
\nonumber\\
\wedge \left. \psi_{unq} \left (  j,i \right )  \right ) 
\Rightarrow \left ( \Psi_{j,i}^{evt} = 1 \right ) 
\end{align}

If any of the conditions in Constraints~(\ref{eq:set-constraint})-(\ref{eq:equiv-constraint}) 
and in Constraint~(\ref{eq:equiv-set-constraint}) is not satisfied between $r_j$ and $r_i$, 
then we do not account for the cache conflict between $r_j$ and $r_i$, as captured by the 
following formulation:
\begin{align}
\label{eq:evict-one-hit-set}
\Theta_{j,i}^{eh} \equiv \left ( \neg \psi_{cnf} \left ( j,i \right ) 
\vee
\neg \psi_{dif} \left ( j,i \right ) \vee \neg \psi_{eqv} \left ( j,i \right ) \right.
\nonumber\\
\vee \neg \left. \psi_{unq} \left ( j,i \right )  \right )
\Rightarrow \left ( \Psi_{j,i}^{evt} = 0 \right ) 
\end{align}

We use variable $miss_i$ to capture whether $r_i$ is a cache miss. 
Therefore, $miss_i$ is set to 1 if $r_i$ is a cache miss, and is set 
to 0 otherwise. We formulate the value of $miss_i$ using the following 
constraints:

\begin{equation}
\label{eq:evict-miss-set-pre}
\Theta_{i}^{mp,lru} \equiv \left ( \sum_{j \in [1,i)} \Psi_{j,i}^{evt} \ge \mathcal{A} \right ) 
\vee \Theta_i^{cold} 
\end{equation}
\begin{equation}
\label{eq:evict-miss-set}
\Theta_{i}^{m,lru} \equiv \left ( \Theta_{i}^{mp,lru} \Rightarrow \left ( miss_i = 1 \right ) 
\right ) 
\end{equation}
\begin{equation}
\label{eq:evict-hit-set}
\Theta_{i}^{h,lru} \equiv \left ( \neg \Theta_{i}^{mp,lru} \Rightarrow \left ( miss_i = 0 
\right ) \right ) 
\end{equation}

In Constraint~(\ref{eq:evict-miss-set-pre}), $\mathcal{A}$ captures the associativity of the cache. 
Once a memory block is loaded into the cache, it requires at least $\mathcal{A}$ unique 
cache conflicts to evict the block. 
If $\Psi_{i,j}^{evt} \ge \mathcal{A}$, $r_i$ has suffered at least $\mathcal{A}$ unique  
cache conflicts since the last access of the memory block referenced by $r_i$ -- resulting 
$r_i$ to be a cache miss. If $r_i$ is not a cold miss ({\em i.e.} $\neg \Theta_i^{cold}$ 
holds) and it has not suffered $\mathcal{A}$ unique cache conflicts, $r_i$ will be a cache 
hit, as captured by Constraint~(\ref{eq:evict-hit-set}).


\subsubsection*{Putting it all together}
To derive the symbolic cache behavior $\Gamma(pc)$, we 
gather all constraints over $\{r_1, \ldots, r_n\}$ and the 
path condition $pc$ as follows: 
{\small
\begin{equation}
\label{eq:gamma-lru-miss-count}
\boxed {
\begin{split}
& \Gamma(pc) \equiv
\\
& pc \wedge \bigwedge_{i \in [1,n]} \left 
( \Theta_i^{m,lru} \wedge \Theta_i^{h,lru} \wedge 
\bigwedge_{j \in [1,i)} \Theta_{j,i}^{em} \right.
\left. \wedge \bigwedge_{j \in [1,i)} \Theta_{j,i}^{eh} \right ) 
\end{split}}
\end{equation}}
$\Theta_i^{m,lru}$ and $\Theta_i^{h,lru}$ together bound the value of 
$miss_i$, which, in turn captures whether $r_i$ is a cache miss. 
However, $\Theta_i^{m,lru}$ and $\Theta_i^{h,lru}$ are dependent on 
symbolic variables $\Psi_{j,i}^{evt}$ where $j \in [1,i)$. The bound 
on $\Psi_{j,i}^{evt}$ is captured via $\Theta_{j,i}^{em}$ and 
$\Theta_{j,i}^{eh}$ 
(Constraints~(\ref{eq:evict-one-miss-set})-(\ref{eq:evict-one-hit-set})). 
Hence, the formulation of $\Gamma(pc)$ includes both 
$\Theta_{j,i}^{em}$ and $\Theta_{j,i}^{eh}$ for $j \in [1,i)$.

\subsubsection*{Complexity of constraints}
The size of our constraint system, in order to check cache side-channel leaks, 
is $O(n^3)$. Here $n$ is the number of memory accesses. The dominating factor 
in our constraint system is the set of constraints generated from 
Constraint~(\ref{eq:evict-one-miss-pre}) and Constraint~(\ref{eq:evict-one-miss-set}). 
In general, we generate constraints for each pair of memory accesses that may potentially 
conflict in the cache, leading to $O(n^2)$ pairs in total. 
For each such pair, the constraint may have a size $O(n)$ 
--- making the size of overall constraint system to be $O(n^3)$. 
However, our evaluation reveals that such a bound is pessimistic and the constraint 
system can be solved efficiently for real-life embedded programs.

\section{Checking Information Leak}
\label{sec:check-information-leak}

In this section, we instantiate \texttt{CHALICE} for two different observer 
models. In particular, we show the formulation of 
Equation~(\ref{eq:heuristic-leak}) by leveraging on our symbolic cache model 
$\Gamma (pc)$ (as described in Sections~\ref{sec:direct-mapped}-\ref{sec:lru-cache}) 
and instantiating $\Phi_{\observer}$ for different observer models. We assume that 
$t_I$ is the observed execution trace for input $I$ and we wish to quantify how much 
information about input $I$ is leaked through $t_I$.



\subsubsection*{Observation via total miss count}

In this scenario, an attacker can observe the number of cache misses in different 
executions~\cite{djb:2005cache}. The observer 
$\mathcal{O}: \Sigma^{*} \rightarrow \mathbb{N}$ is a function, where a sequence 
of cache hits and misses are mapped to a non-negative integer capturing the number 
of cache misses. 
Therefore, for a given trace $t \in \Sigma^{*}$, $\mathcal{O}(t)$ captures the number 
of cache misses in the trace $t$.

Recall that we use variable $miss_i$ to capture whether the $i$-th memory access 
was a cache miss. We check the unsatisfiability of the following logical formula to 
record information leak:
%
\begin{equation}
\label{eq:information-leak-count-1}
\bigvee_{e \in \mathit{Path}} \left ( \Gamma(pc_e) \wedge 
\left ( \displaystyle \sum_{i \in [1,n_e]} miss_{i} = \mathcal{O}(t_I) \right ) 
\wedge \pi \right )
\end{equation}
where $n_e$ is the number of memory accesses occurring along path $e$ and $\pi$ is a 
predicate defined on program inputs. 
Concretely, if Constraint~(\ref{eq:information-leak-count-1}) is unsatisfiable, we can 
establish that the information ``$\neg \pi \equiv true$" is leaked through the execution 
trace $t_I$. By performing such unsatisfiability checks over the entire program input 
space, we quantify the information leak $\leak (t_I)$ through execution trace $t_I$ 
({\em cf.} Equation~(\ref{eq:heuristic-leak-bound})).    


\subsubsection*{Observation via hit/miss sequence}
%
For an execution trace $t \in \Sigma^{*}$, an observer can monitor hit/miss sequences 
from $t$~\cite{trace-attack-paper}. Concretely, let us assume $\{o_1, o_2, \ldots, o_k\}$ 
is the set of positions in trace $t$ where the observation occurs. If $n$ is the total 
number of memory accesses in $t$, we have $o_i \in [1,n]$ for each $i \in [1,k]$.  

We define the observer $\mathcal{O}: \Sigma^{*} \rightarrow \{0,1\}^{k}$ as a 
projection from the execution trace onto a bitvector of size $k$. Such a 
projection satisfies the following conditions:  
${\mathcal{O}(t)}_i = 1$ if $t_{o_i} = m$ and ${\mathcal{O}(t)}_i = 0$ otherwise. 
${\mathcal{O}(t)}_i$ captures the $i$-th bit of $\mathcal{O}(t)$ and 
similarly, $t_{o_i}$ captures the $o_i$-th element in the execution trace $t$. 
Note that a strong observer could map the entire execution 
trace to a bitvector of size $n$.

For such an observer, we check the unsatisfiability of the following formula to 
record information leak: 
{\small
\begin{equation}
\label{eq:information-leak-sequence-1}
\bigvee_{e \in \mathit{Path}} \left ( \Gamma \left ( pc_e \right ) \wedge 
\bigwedge_{i \in \{1,2,\ldots,k\}} 
\left (
    \begin{array}[b]{c}
      o_i \leq n_e\\
      \wedge miss_{o_i} = {\mathcal{O}(t_I)}_i 
    \end{array} 
\right ) 
\wedge \pi \right )
\end{equation}}
where $\pi$ is a predicate on program inputs. By generating such predicates over 
the input space, we quantify the information leaked about input $I$ via $\leak (t_I)$ 
({\em cf.} Equation~(\ref{eq:heuristic-leak-bound})).



Although we instantiate \texttt{CHALICE} for two observer models, we 
believe that our framework is generic to capture a wide range of such models. 
In particular, we can tune \texttt{CHALICE} for any observer model that is expressed 
via symbolic constraints over variables $miss_i$.

\section{Implementation Aspects}
\label{sec:implementation}
%

In this section, we discuss some crucial implementation aspects 
for the efficiency and effectiveness of \texttt{CHALICE}.  
 
\subsubsection*{Implementation setup}
We implemented \texttt{CHALICE} on top of the KLEE symbolic virtual 
machine~\cite{klee-url}. However, in order to design such an 
implementation, we faced the following challenges. 

\begin{figure}[h]
\rotatebox{0}{
\includegraphics[scale = 0.3]{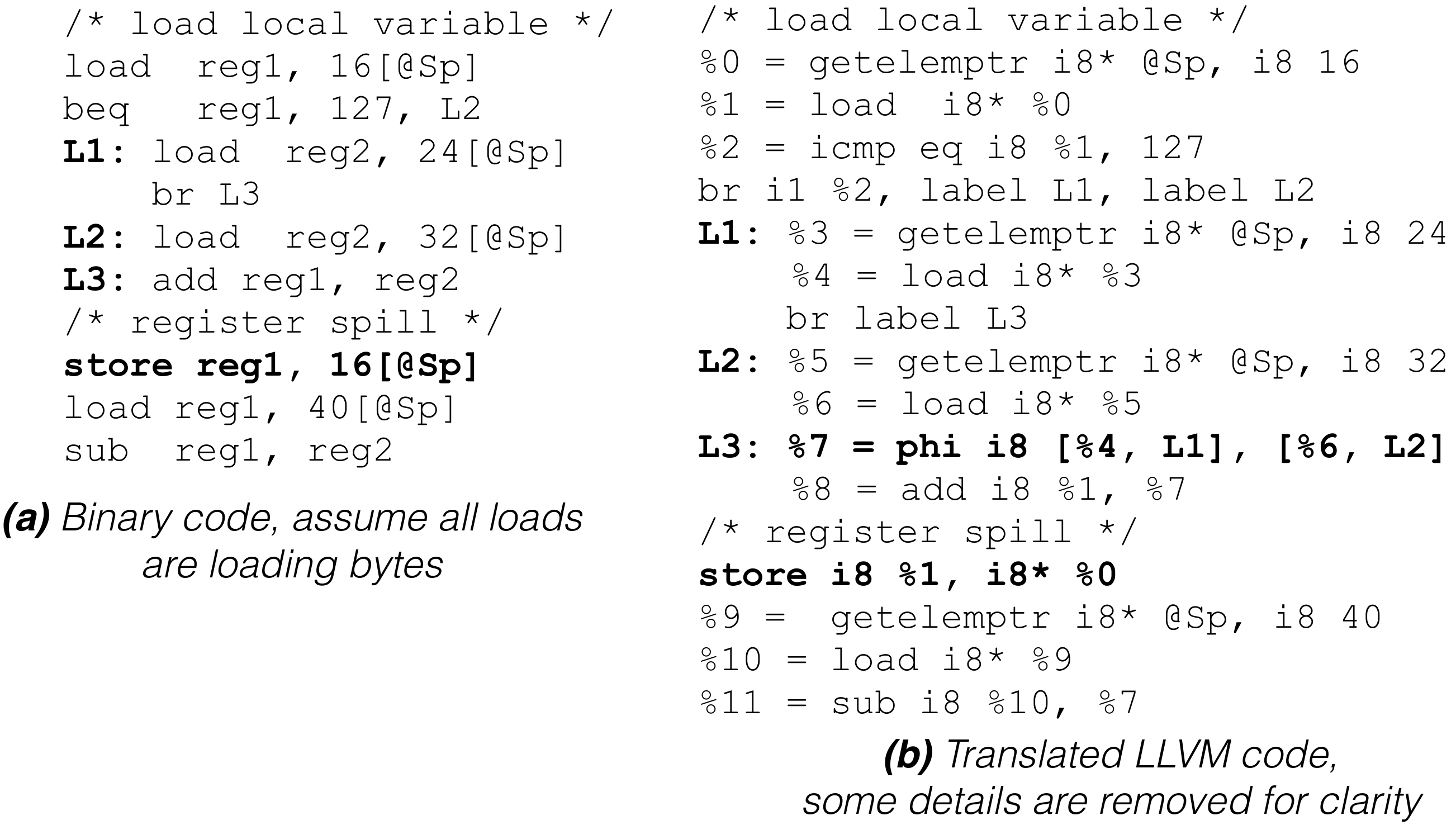}}
\caption{Translation from binary code to LLVM code}
\label{fig:binary-example}
\end{figure}

KLEE works on LLVM bitcode~\cite{llvm-url}. Considering cache 
performance, at the level of LLVM bitcode, introduces several 
inaccuracies. For instance, LLVM bitcode uses an unbounded 
number of virtual registers. In contrast, any given execution 
platform only contains a finite number of physical registers. 
In order to understand how this impacts memory performance, 
consider the example in Figure~\ref{fig:binary-example}.

In Figure~\ref{fig:binary-example}, we assume that the execution 
platform contains only two physical registers. As a result, a 
register spill is required in the binary code to preserve the 
functionality of the LLVM bitcode. 
In general, aggressive compiler optimizations may change the 
structure and memory behaviour of the LLVM bitcode dramatically, 
when translated into native binary. 

In order to solve this challenge and still use the power of 
symbolic execution on target-independent LLVM bitcode, we 
have designed a translator that converts binary code to LLVM 
bitcode. Such a translation must preserve the following properties 
to produce a valid LLVM bitcode.  
First, we ensure that each load/store instruction in the binary code 
to have a functionally equivalent load/store instruction in the 
translated bitcode. Secondly, we preserve the static-single-assignment 
(SSA) form of LLVM bitcode by systematically inserting $\mathit{Phi}$ 
functions. Thirdly, several instructions at the machine code may 
require multiple LLVM instructions to implement. The \texttt{LWL} 
and \texttt{LWR} are such machine-level instructions for MIPS 
architecture.   
Finally, LLVM bitcode is strongly typed. As a result, 
LLVM bitcode uses different instructions for pointer arithmetic 
as compared to general-purpose arithmetic. We use a lightweight 
type inference on the binary code and compute the appropriate 
LLVM instruction for a given machine-level instruction. 
Figure~\ref{fig:binary-example}(b) demonstrates how the example 
binary code is translated into LLVM bitcode. The instruction 
\texttt{getelemptr} handles pointer arithmetic in the LLVM 
bitcode. 

From a technical point of view, we have designed a translator that 
converts PISA binaries (a MIPS like architecture) into LLVM bitcode. 
Such a translator is unique in the sense that it focuses on preserving 
the memory behaviour during the translation. 
Nevertheless, our translator may introduce additional instructions 
to preserve the SSA semantics of LLVM bitcode. Such additional 
instructions are not part of the binary code. In order to exclude 
such instructions from our analysis, we annotate the LLVM bitcode 
with a mapping from each memory-related instruction in the binary 
to the respective memory-related instruction in the LLVM bitcode.  
As a result, \texttt{CHALICE} accurately captures the cache 
side-channel leaks for applications compiled into PISA binaries. 

Our translator currently does not handle indirect jump instructions. 
However, we can use a lightweight static analysis to compute the 
potential targets for indirect jumps and the translator can easily 
be modified to take this into account. Besides, \texttt{CHALICE} 
is modular in the sense that it can easily be adapted for a different 
architecture. This can be accomplished only by extending the 
translator to convert the respective machine code into LLVM bitcode.


\subsubsection*{Reducing the number of constraints}
In order to reduce the size of $\Gamma(pc)$, we first inspect constraints 
generated for each memory-related instruction individually. 
In particular, for each memory-related instruction $r_i$, we check whether 
the respective memory access leads to a cache miss (or hit) for all inputs 
satisfying $pc$. For instance, consider 
Constraints~(\ref{eq:evict-one-miss})-(\ref{eq:evict-one-hit}) for direct-mapped 
caches. In order to check whether instruction $r_i$ is a miss 
for all inputs $I \Rightarrow pc$, we check the validity of the constraint 
$pc \wedge \Theta_{i}^{mp,dir}$. Similarly, we check the unsatisfiability of the 
constraint $pc \wedge \Theta_{i}^{mp,dir}$, to prove that $r_i$ is always a cache 
hit for all inputs $I \Rightarrow pc$. If $pc \wedge \Theta_{i}^{mp,dir}$ 
is valid (resp. unsatisfiable), we can directly consider $miss_i$ to 
be 1 (resp. 0) within the symbolic cache model $\Gamma (pc)$. As a result, we 
discard all constraints $\Theta_{i}^{m,dir}$ and $\Theta_{i}^{h,dir}$ 
in formulating $\Gamma(pc)$. It is worthwhile to note that this optimization 
increases the time to process a single memory-related instruction, as the 
solver is called at each memory access. However, we discovered that in 
practice, this step dramatically reduces the size of $\Gamma(pc)$, making 
our information leak detection tractable.

\section{Evaluation}
\label{sec:evaluation}

\paragraph*{\textbf{Experimental setup}}


In order to evaluate the effectiveness of \texttt{CHALICE}, we have chosen 
cryptographic applications from OpenSSL library~\cite{openssl-url} and other 
software repository~\cite{aes-impl}, as well as applications from Linux GDK 
library. The choice of our subject programs is motivated by the critical 
importance of validating security-related properties in these applications. 
Some salient features of the evaluated subject programs is outlined in 
Table~\ref{tab:case-study}. We have performed all our experiments on an Intel 
I7 machine with 8GB of RAM and running Debian as operating systems.     


\begin{table}[t]
  \begin{center}
  {\scriptsize 
  \begin{tabular}{|c||c|c|c|}
  \hline
   Program & Lines of  & Lines of & Max. \#Memory \\
								& C code & LLVM code & access\\
    \hline
      \texttt{AES}~\cite{aes-impl} & 800 & 4950 & 2134\\
      \hline
      \texttt{AES}~\cite{openssl-url} & 1428 & 1800 & 420\\
      \hline
      \texttt{DES}~\cite{openssl-url} & 552 & 3990 & 334\\
      \hline
      \texttt{RC4}~\cite{openssl-url} & 160 & 668 & 1538\\
      \hline
      \texttt{RC5}~\cite{openssl-url} & 256 & 1820 & 410\\
      \hline
      \hline
      \texttt{gdk\_keyval\_to\_unicode} & 1300 & 268 & 114\\
      \hline
    	  \texttt{gdk\_keyval\_name} & 1350 & 1408 & 12\\
	  \hline
    \end{tabular}}
  \vspace*{0.05in}  
  \caption{Salient features of the evaluated subject programs}
  \vspace*{-0.3in}
  \label{tab:case-study}   
\end{center}
\end{table}


\subsection{\textbf{Generating Predicates on Inputs}}
\label{sec:evaluation-input-predicates}
Using \texttt{CHALICE}, we can select an arbitrary number of bits in the 
program input to be symbolic. These symbolic bits capture the high 
sensitivity of the input subspace and our framework focuses to quantify 
the information leaked about this subspace. For instance, in encryption 
routines, the bits of private input ({\em e.g.} a secret key) can be made 
symbolic. Without loss of generality, in the following discussion, we 
assume that the entire input is sensitive and we make all input bits to 
be symbolic. 

%
Let us assume $N$-byte program input. We use the notation $k[i]$ to capture 
the $i$-th byte of an arbitrary input $k$. Similarly, we use $k[i,j]$ to 
capture the $j$-th bit of the $i$-th byte in $k$.  
We generate the following predicates on inputs for quantifying information 
leak $\leak (t_I)$ ({\em cf.} Equation~(\ref{eq:heuristic-leak-bound})).   
\begin{equation*}
\mathbf{P_{bit}} = \left \{ k[i,j]=v\ |\ i \in [1,N], j \in [1,8], v \in [0,1] \right \}
\end{equation*}
\begin{equation*}
\mathbf{P_{byte}} = \left \{ k[i]=v\ |\ i \in [1,N], v \in [0,255] \right \}
\end{equation*}

It is worthwhile to mention that for a 16-byte sensitive input ({\em e.g.} in AES-128), 
$\mathbf{P_{bit}}$ and $\mathbf{P_{byte}}$ lead to 256 and 4096 calls to the solver, 
respectively to quantify $\leak (t_I)$. 

%
%
%

\begin{figure*}[t]
\begin{center}
\begin{tabular}{cc}
\rotatebox{0}{
\includegraphics[scale = 0.33]{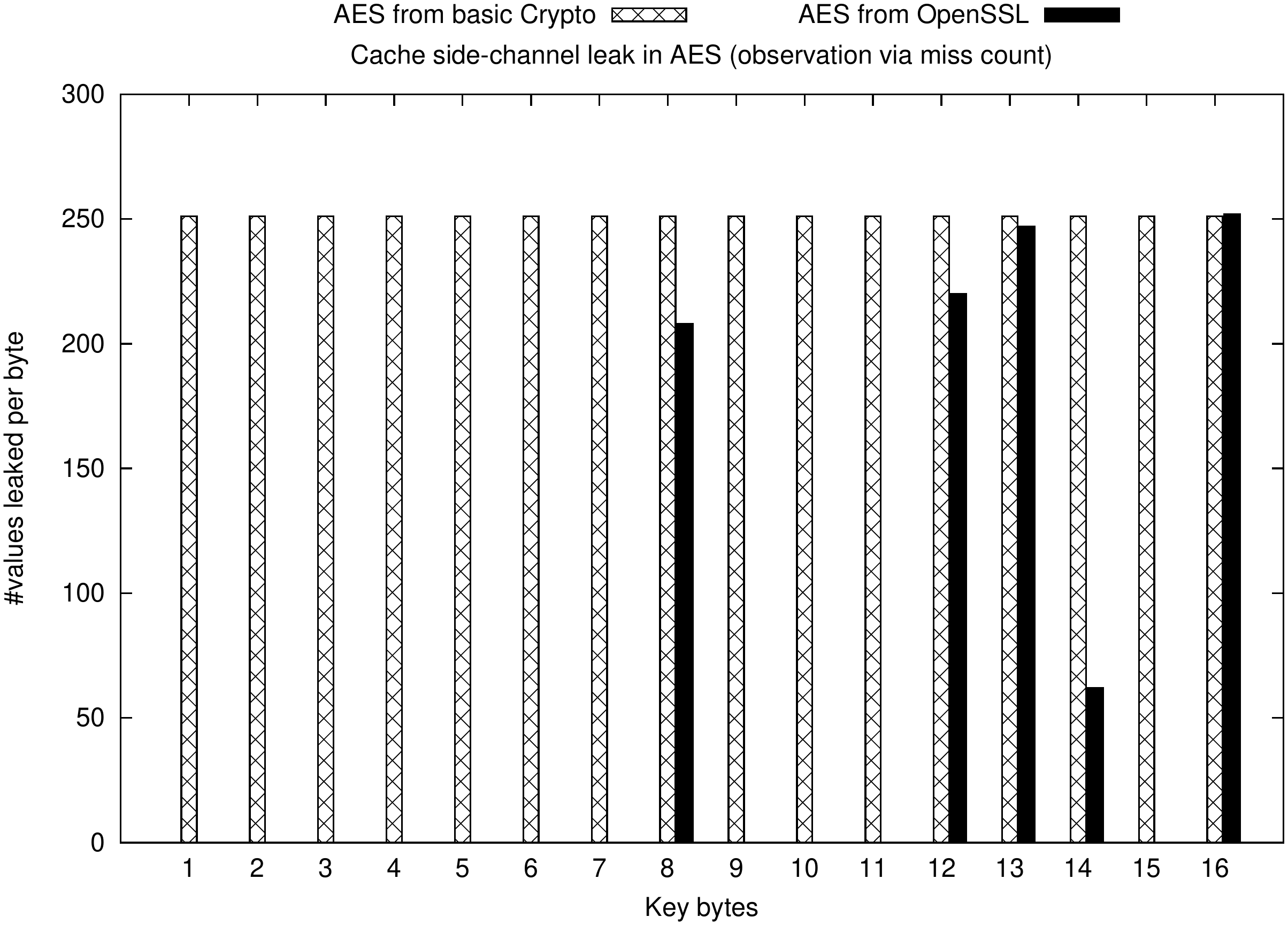}} &
\rotatebox{0}{
\includegraphics[scale = 0.33]{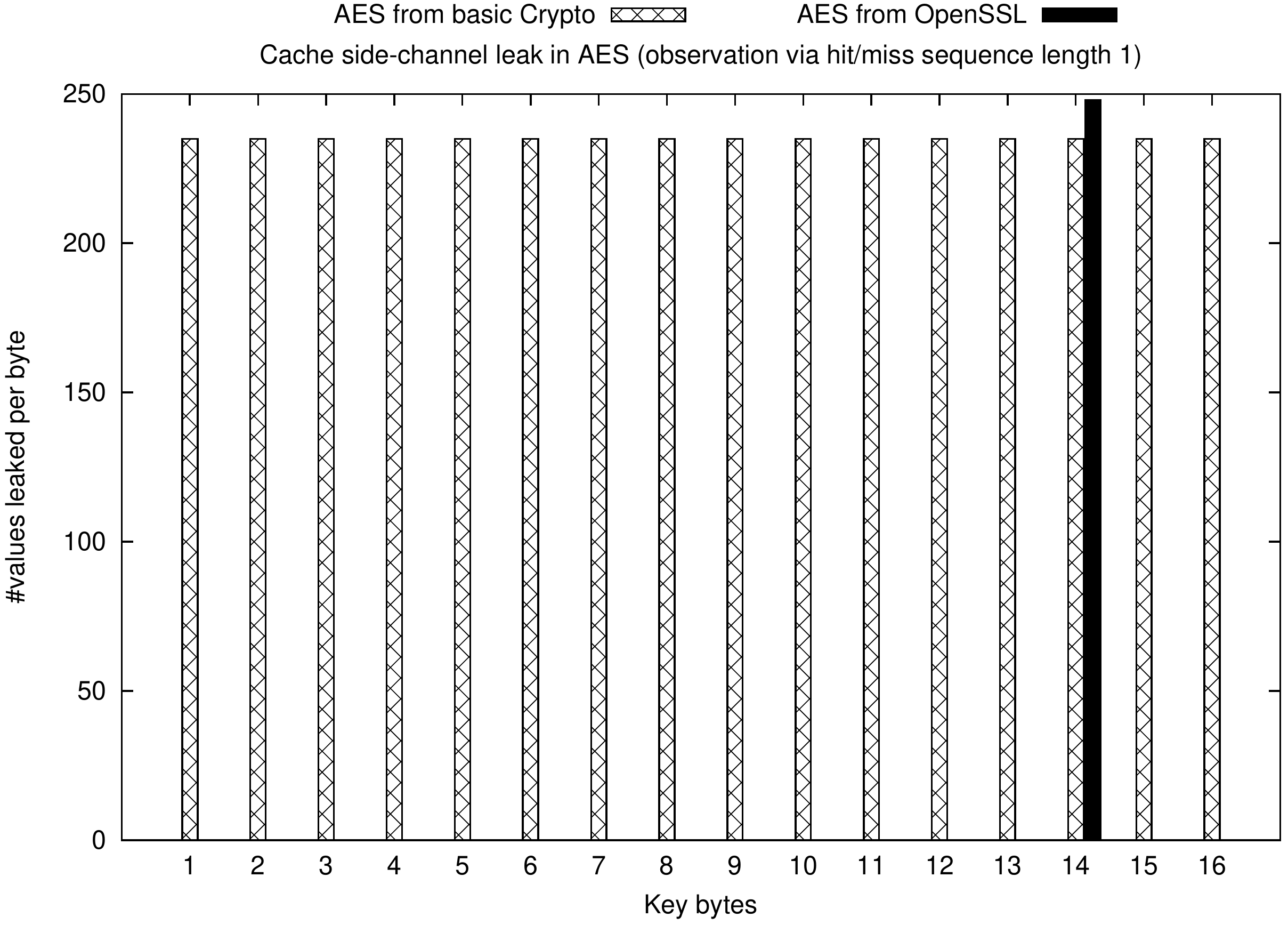}}\\
\textbf{(a)} & \textbf{(b)}
\end{tabular}
\end{center}
\vspace*{-0.1in}
\caption{Information leak observed using an 8 KB direct-mapped cache}
\label{fig:cache-miss-byte}
\end{figure*}

\begin{figure*}[t]
\begin{center}
\begin{tabular}{cc}
\rotatebox{0}{
\includegraphics[scale = 0.33]{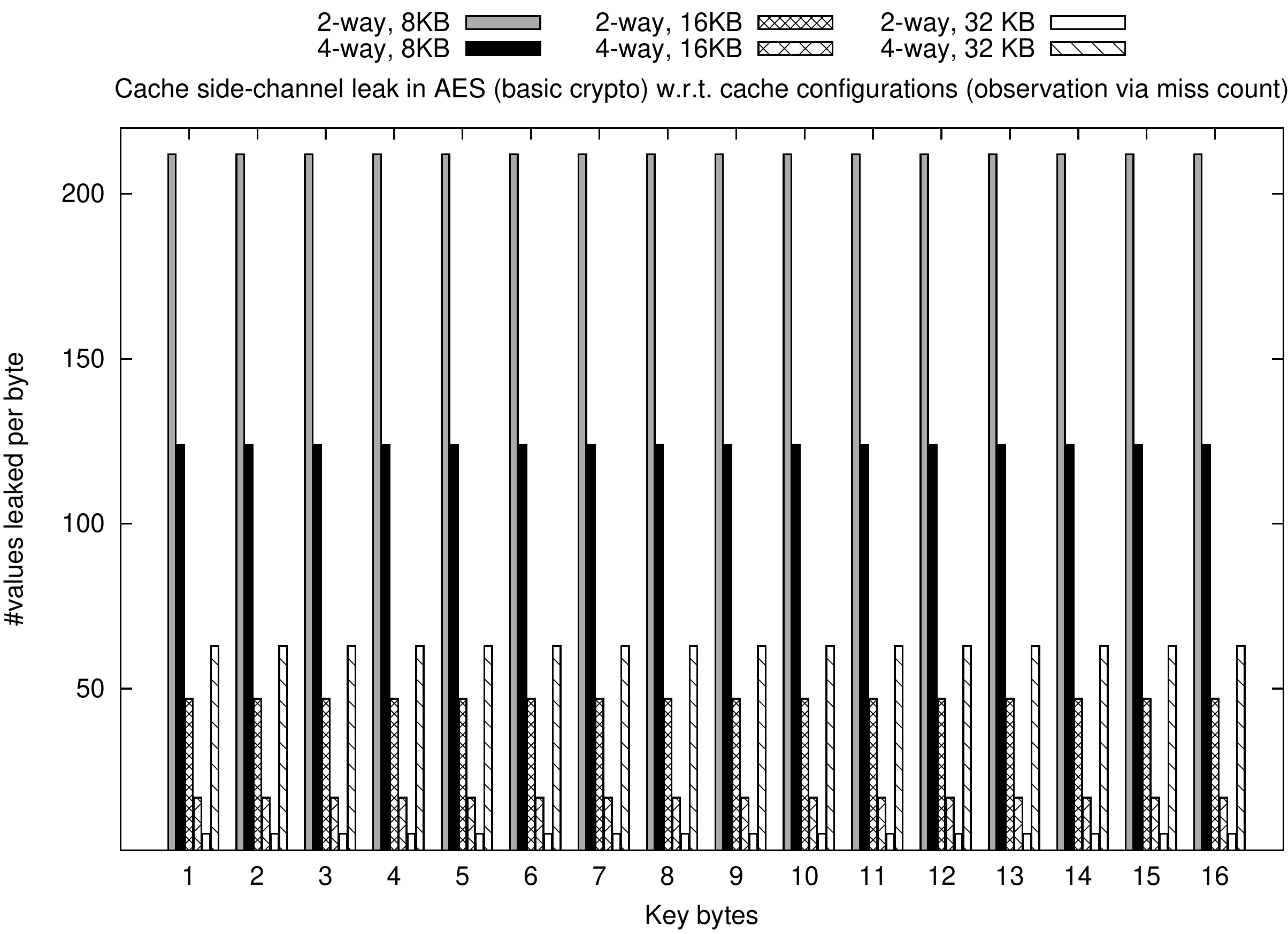}} &
\rotatebox{0}{
\includegraphics[scale = 0.33]{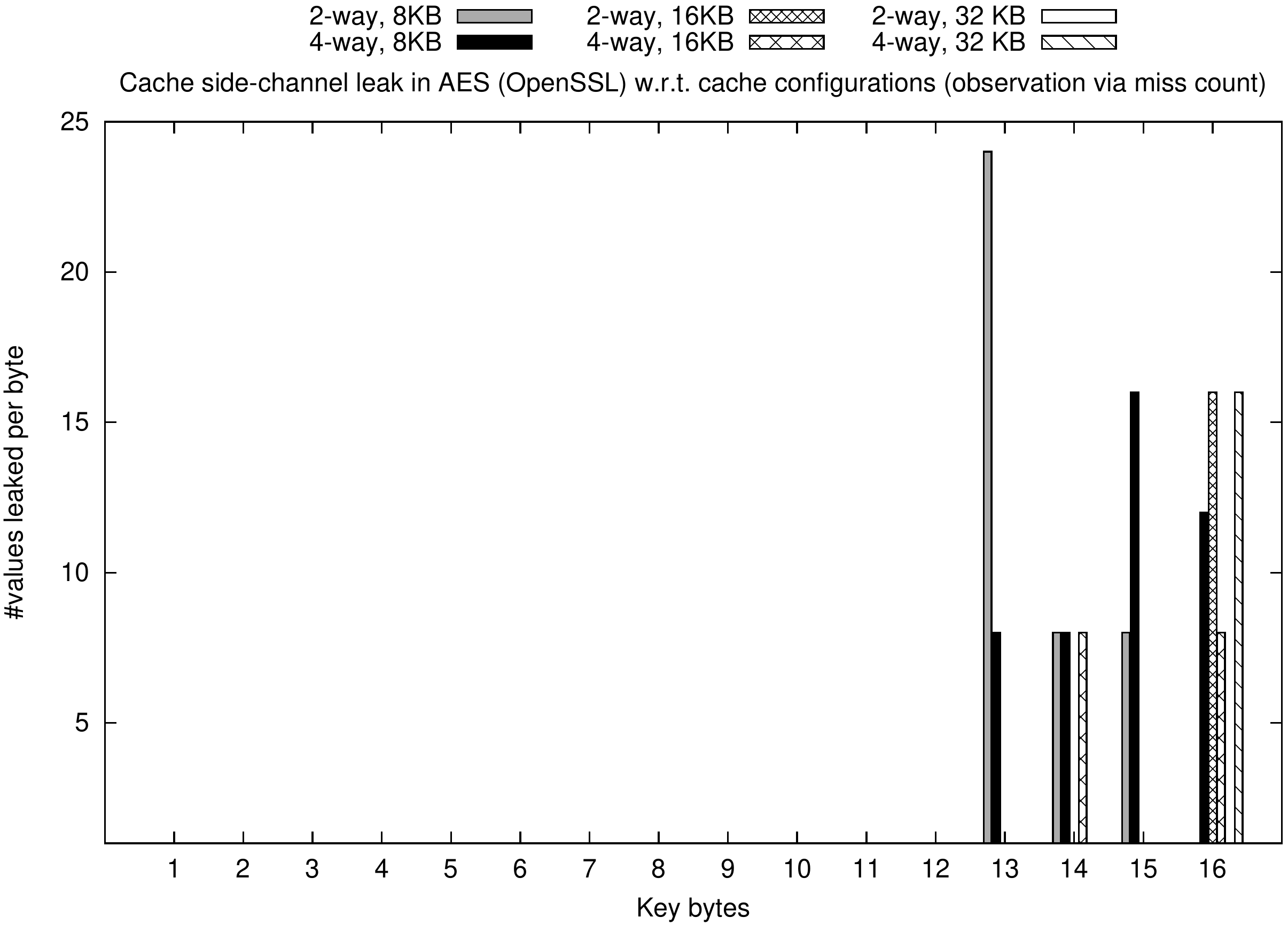}}\\
\textbf{(a)} & \textbf{(b)}
\end{tabular}
\end{center}
\vspace*{-0.1in}
\caption{Information leak in AES w.r.t. observations based on miss count (set-associative caches employing LRU policy)}
\label{fig:cache-miss-byte-sens}
\end{figure*}

\subsection{Experience with AES-128}
\label{sec:evaluation-aes}
We used two different implementations~\cite{openssl-url,aes-impl} of 
the Advanced Encryption Standard (AES).
AES is a widely used encryption standard for achieving confidential 
communication. AES has been of great importance for delivering security 
in embedded systems because of its sound protection strength and high 
throughput ({\em e.g.} even on credit cards). Therefore, it is crucial 
to validate security-related properties, such as side-channel resistance, 
for AES. 

AES has input-dependent memory accesses. In particular, different encryption 
rounds of AES revolve around accessing an \texttt{sbox} -- a matrix-like 
structure kept in main memory (DRAM). During encryption, AES code accesses 
varied locations in the \texttt{sbox}. The location of the \texttt{sbox} being 
accessed, for a given instruction, depends on the secret key. 
That is, the sequence of memory blocks, accessed during encryption, is dependent 
on the value of secret key. As a result, we potentially obtain different cache 
performance for different secret keys. 

\subsubsection{Key result}

\begin{figure*}[t]
\begin{center}
\begin{tabular}{cc}
\rotatebox{0}{
\includegraphics[scale = 0.33]{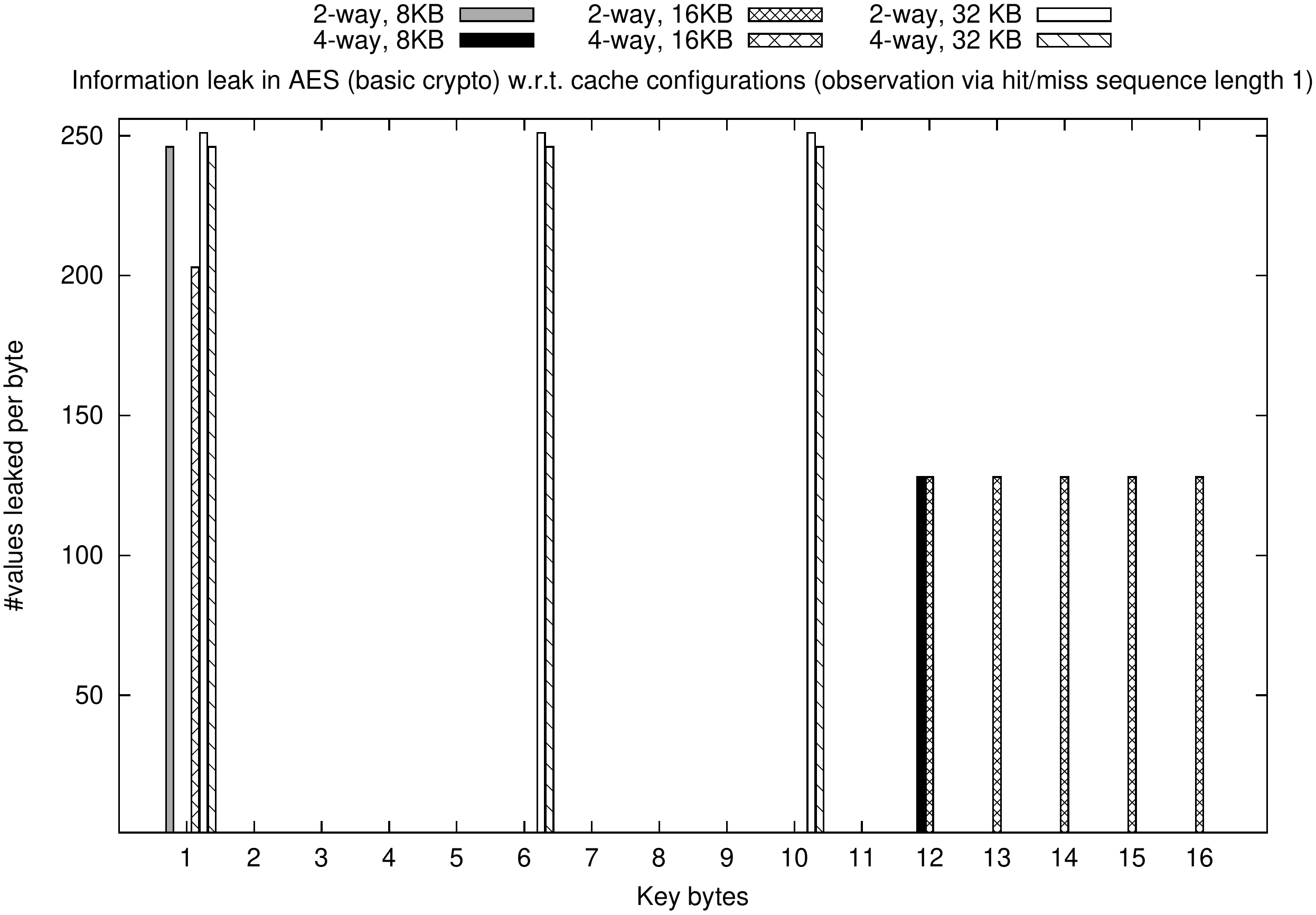}} &
\rotatebox{0}{
\includegraphics[scale = 0.33]{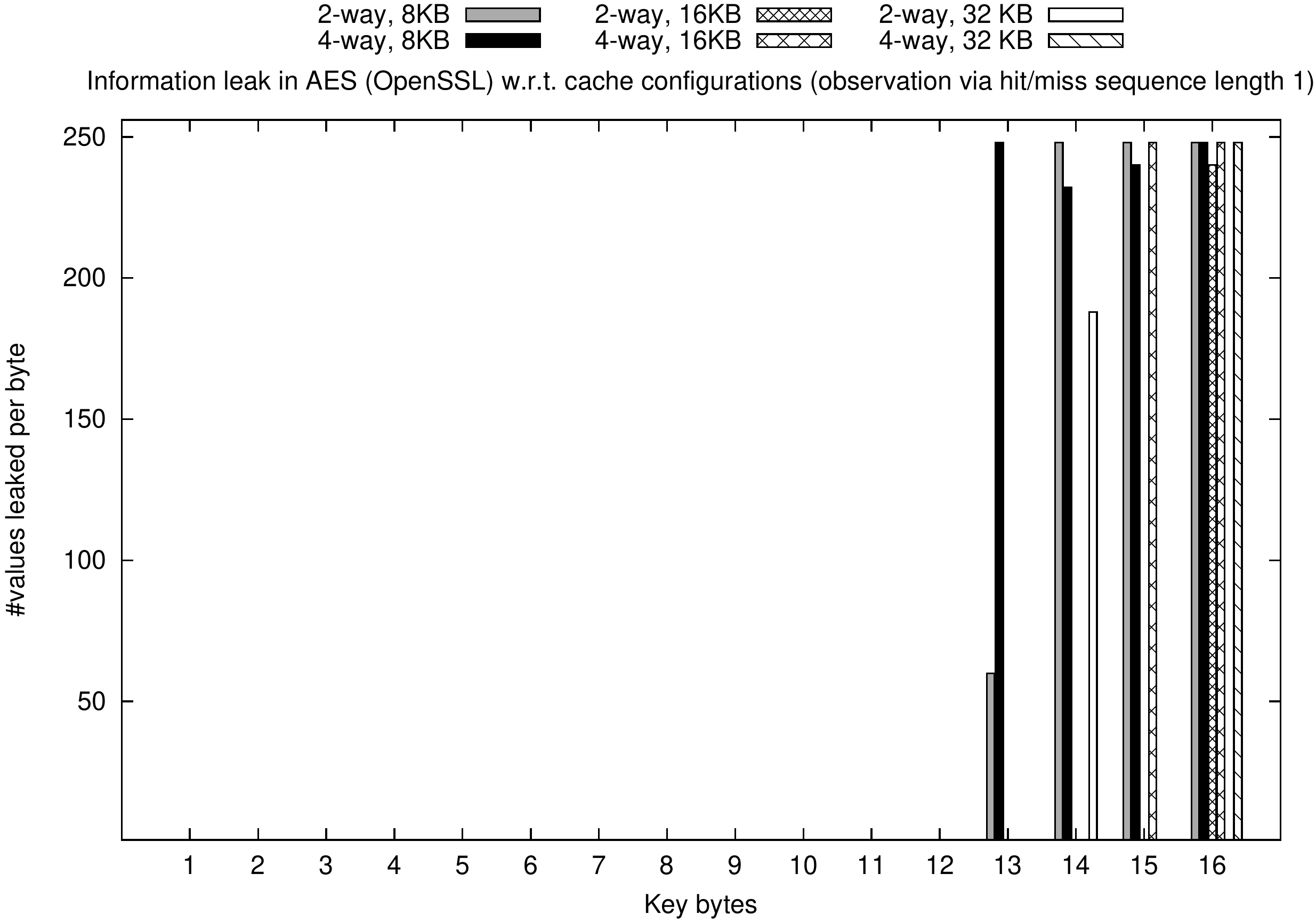}}\\
\textbf{(a)} & \textbf{(b)}
\end{tabular}
\end{center}
\vspace*{-0.1in}
\caption{Information leak in AES w.r.t. observations based on hit/miss sequence}
\label{fig:cache-miss-byte-seq-sens}
\end{figure*}

We used an 8 KB direct-mapped cache with a line size of 32 bytes. 
This size is big enough to keep the entire \texttt{sbox} of AES 
in the cache. 
%
We executed AES in simplescalar simulator \cite{simplescalar-tool} 
({\em cf.} Figure~\ref{fig:aes-gaussian}) with a test suite and obtained 
the respective set of observations ({\em e.g.} number of cache misses). 
For such observations, we intended to check how much information is leaked 
through a bit or a byte, by generating predicates $P_{bit}$ and $P_{byte}$ 
(as described in Section~\ref{sec:evaluation-input-predicates}), respectively. 


For the collected set of observations, \texttt{CHALICE} quantifies $\leak (t_I)$ 
to be 0 when the set of predicates $P_{bit}$ is used. Therefore, each observation 
({\em e.g.} the number of cache misses) is possible irrespective of whether an 
arbitrary bit of the AES input (in both implementations of AES~\cite{aes-impl,openssl-url}) 
is ``1" or ``0". Therefore we deduce, for the given set of observations, 
{\em there does not exist any dependency between the cache performance 
and the value of an arbitrary bit of the key.}

Figure~\ref{fig:cache-miss-byte}(a) captures an outline of information leak 
highlighted via \texttt{CHALICE}, for two different implementations of 
AES~\cite{aes-impl,openssl-url}. For each byte of the 16-byte secret key, 
we show the amount of information leaked through the number of cache misses. 
For instance, we establish for certain observations, that as many as 251 values 
(out of 256) are leaked for each byte of AES key (in the implementation~\cite{aes-impl}). 
{\em This means, there exists at least $251^{16}$ possible keys 
(out of a total $2^{128}$) that can be eliminated just by observing the 
cache misses}. Such an information gives the designer valuable insights when 
designing embedded systems, both in terms of choosing an AES key and a cache 
architecture, in order to avoid serious security breaches.  
In contrast to the implementation of~\cite{aes-impl}, we can observe from 
Figure~\ref{fig:cache-miss-byte}(a) that the implementation of AES from OpenSSL 
exhibits substantially fewer information leaks. For instance, certain key bytes 
of OpenSSL AES do not leak any information through the number of cache misses. 

In our framework, we also investigated on adversaries who can observe 
the sequence of cache hits and misses, instead of just the overall 
number of cache misses. However, to simplify our evaluation, we 
focused on sequences of length 1, and considered all the memory 
accesses. Our goal is to check the dependency between the AES-key and 
the hit/miss characteristics of an arbitrary memory access.

Figure~\ref{fig:cache-miss-byte}(b) captures a snapshot of dependencies 
between AES-key bytes and the cache behavior of different memory accesses. 
For instance, the maximum values leaked through a byte can be as high as 235, 
as shown via Figure~\ref{fig:cache-miss-byte}(b). Similar to 
Figure~\ref{fig:cache-miss-byte}(a), we also observe that the AES implementation 
from OpenSSL leaks substantially less information, as compared to the implementation 
in~\cite{aes-impl}, when cache behavior is observed individually for each memory access.  




\subsubsection{Sensitivity of Information Leak w.r.t Cache Size}
Figures~\ref{fig:cache-miss-byte-sens}-\ref{fig:cache-miss-byte-seq-sens} 
capture the sensitivity of information leakage with respect to different 
configurations. For all experiments, the replacement policy is set to LRU 
and the cache-line size is set to a fixed 32 bytes. 
Figures~\ref{fig:cache-miss-byte-sens}(a)-(b) captures the 
information-leakage-sensitivity for observations via a given number of 
cache misses.
Increasing cache size (or associativity) may have two contrasting effects 
as follows. For a given cache size, let us assume a subset of the input 
space $\mathbb{I}_{= C} \subseteq \mathbb{I}_{< C} \cup \mathbb{I}_{= C} \cup \mathbb{I}_{> C}$ 
(where $\mathbb{I}_{< C} \cup \mathbb{I}_{= C} \cup \mathbb{I}_{> C}$ is the 
entire input space) which leads to $C$ cache misses. Increasing cache size 
reduces cache conflict. Therefore, it is possible that some input 
$i \in \mathbb{I}_{> C}$, which 
leads to more than $C$ cache misses with a smaller cache, produces 
$C$ cache misses with the increased cache size. This tends to increase the
number of inputs leading to $C$ cache misses, thus reducing the amount of 
information leaked through observing $C$ misses. Secondly, some input 
$i \in \mathbb{I}_{= C}$ may have less than $C$ cache misses with increased 
cache size. This may reduce the number of inputs having $C$ cache misses, 
thus increasing the potential leakage through the observation of $C$ cache 
misses. In Figure~\ref{fig:cache-miss-byte-sens}(a), the reduction in cache 
side-channel leakage is visible for cache sizes up to 16 KB, for AES 
implementation from~\cite{aes-impl}. However, for a 4-way 32 KB cache, 
we observe the increase in information leakage. This is 
because the number of possible keys, leading to a given observation, 
is reduced considerably.  

Figures~\ref{fig:cache-miss-byte-seq-sens}(a)-(b) capture the sensitivity 
of information leakage (w.r.t. cache size) for an adversary who can observe 
the cache behavior of an arbitrary memory access. 
Concretely, consider the bars in Figure~\ref{fig:cache-miss-byte-seq-sens}(a) 
for 8 KB and 32 KB caches. For a 2-way, 8 KB cache, a significant information 
about the first key byte is leaked.
With 32 KB caches, the number of cache conflicts reduces substantially, but 
we observe substantial leakage of information about key bytes one, six and 
ten. Therefore, even though the increased cache size improves performance, 
it might make the overall system potentially less secure, as shown in 
Figure~\ref{fig:cache-miss-byte-seq-sens}(b). In summary, we believe such 
insights are valuable for designers to build secure systems.

The evaluation also reveals that the AES implementation from OpenSSL exhibits 
information leak, as shown in Figure~\ref{fig:cache-miss-byte-seq-sens}(b). 
Figure~\ref{fig:cache-miss-byte-seq-sens}(b) highlights the last four bytes of 
the key to experience more leakage of information as compared 
to other key bytes. 



\begin{figure*}[t]
\begin{center}
\begin{tabular}{cc}
\rotatebox{0}{
\includegraphics[scale = 0.33]{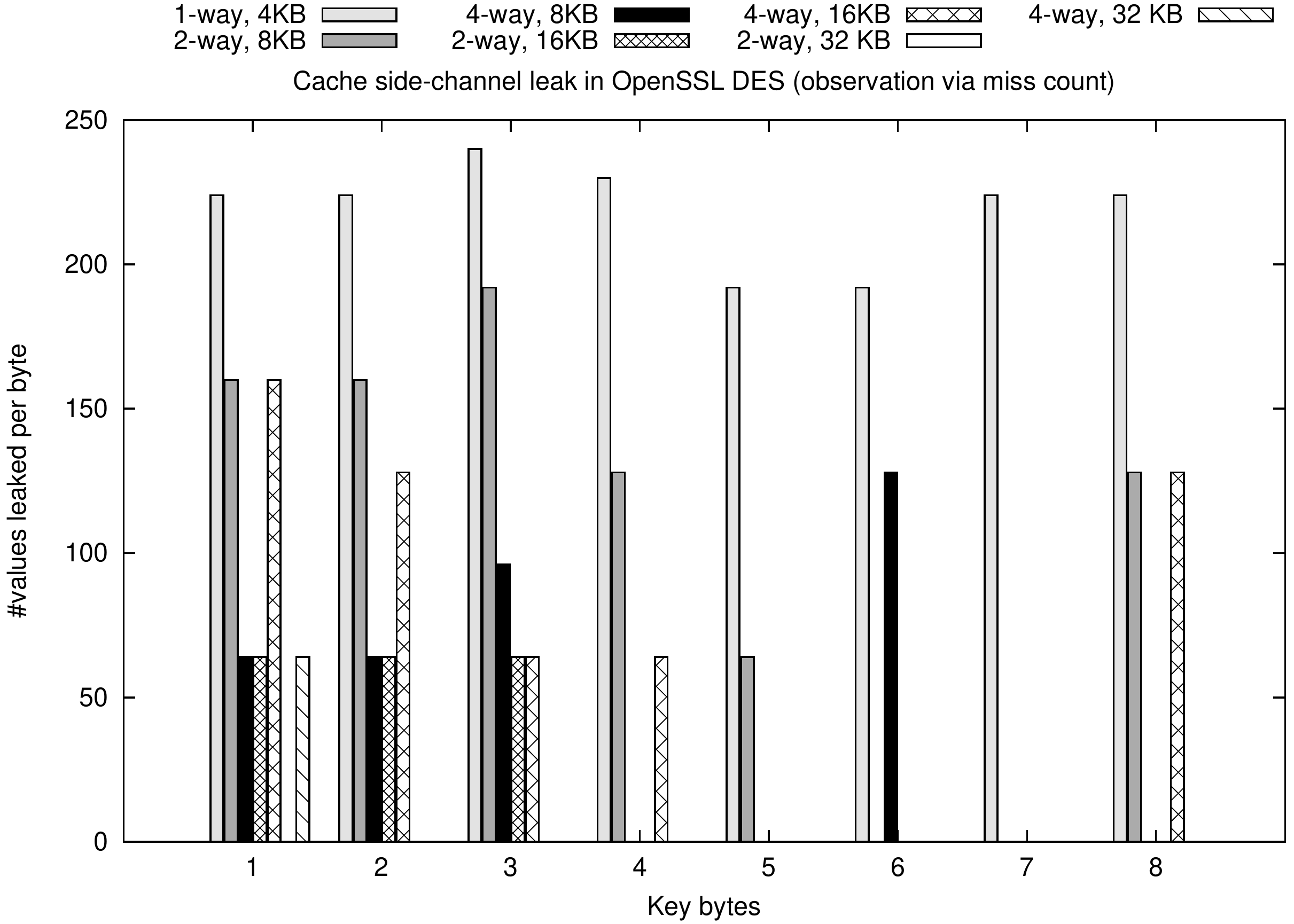}} &
\rotatebox{0}{
\includegraphics[scale = 0.33]{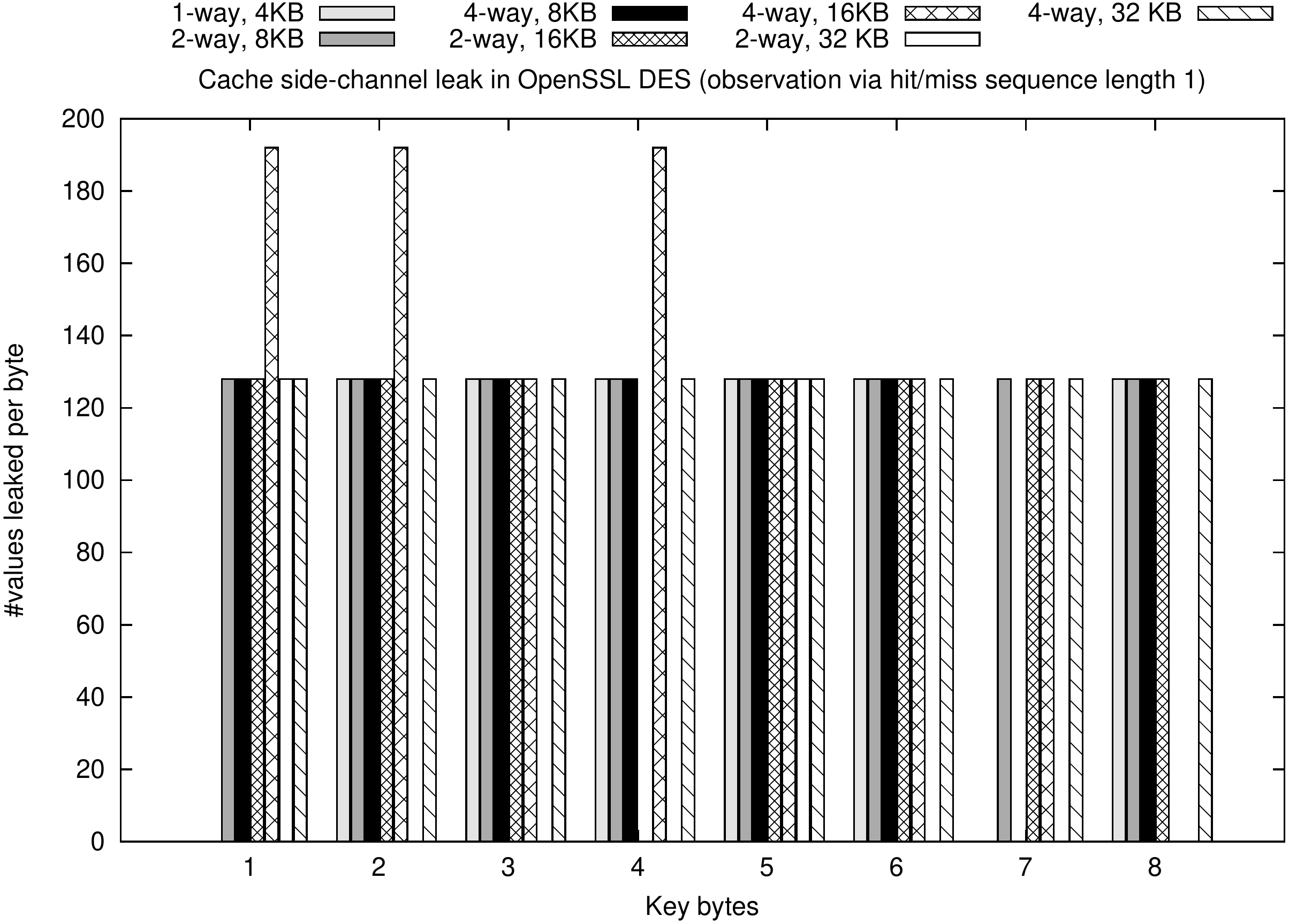}}\\
\textbf{(a)} & \textbf{(b)}
\end{tabular}
\end{center}
\vspace*{-0.1in}
\caption{Information leak in DES w.r.t. different observer models}
\label{fig:des-result-sens}
\end{figure*}

\subsection{Experience with DES}
\label{sec:evaluation-des}
Data Encryption standard (DES)~\cite{openssl-url} is a symmetric key algorithm for 
electronic data. 
%
The OpenSSL implementation of DES encrypts 64 bit message with a 64 bit secret 
key. 
Figures~\ref{fig:des-result-sens}(a)-(b) summarize our result on quantifying 
information leak in DES. Figure~\ref{fig:des-result-sens}(a) reports information 
leaks through observing miss count. 
For instance, using 8KB caches, DES leaks more than 150 values for several key bytes. 
In contrast, information leak in the OpenSSL version of AES is relatively sparse and 
it generally leaks less information about key bytes 
({\em cf.} Figure~\ref{fig:cache-miss-byte-sens}(b)). 
%
In Figure~\ref{fig:des-result-sens}(b), we observe a similar trend, as DES continues 
to suffer from information leak when the cache behavior of an arbitrary memory access 
is observed. Our results summarize potentially insecure nature of DES, even if we only 
consider security leaks through cache behaviour.

\begin{figure*}[t]
\begin{center}
\begin{tabular}{cc}
\rotatebox{0}{
\includegraphics[scale = 0.33]{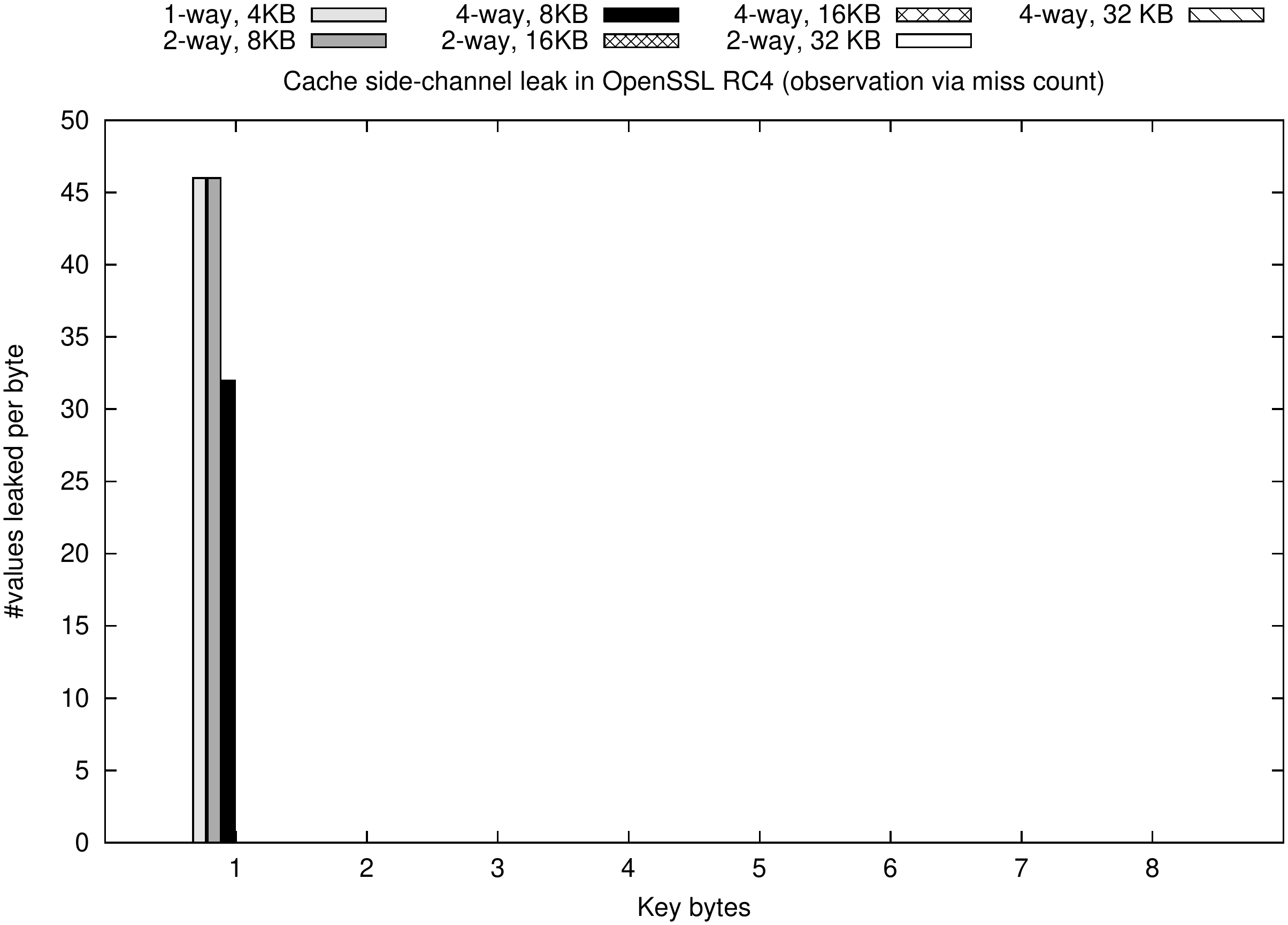}} &
\rotatebox{0}{
\includegraphics[scale = 0.33]{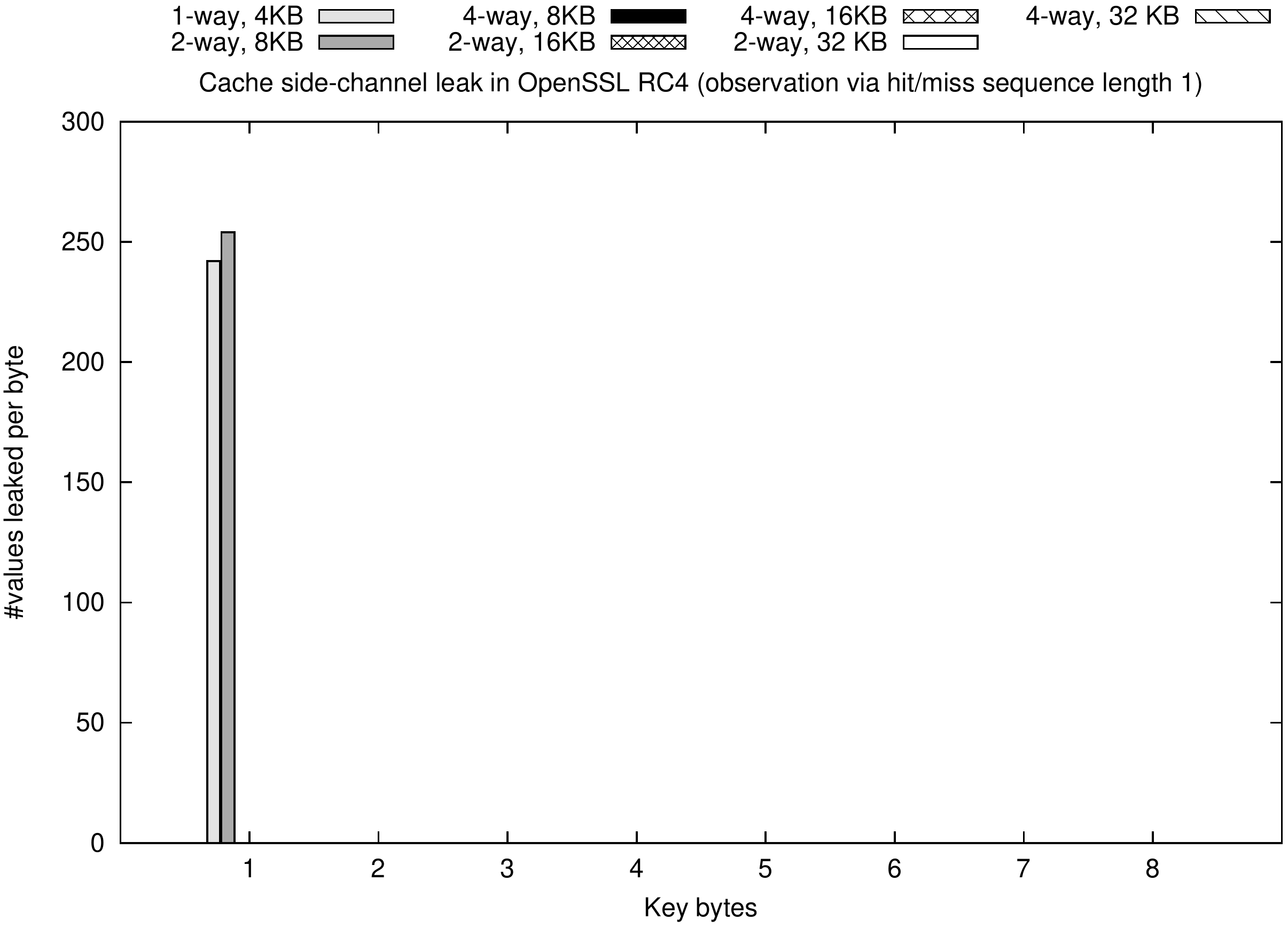}}\\
\textbf{(a)} & \textbf{(b)}
\end{tabular}
\end{center}
\vspace*{-0.1in}
\caption{Information leak in RC4 w.r.t. different observer models}
\label{fig:rc4-result-sens}
\end{figure*}

\begin{figure*}[t]
\begin{center}
\begin{tabular}{cc}
\rotatebox{0}{
\includegraphics[scale = 0.33]{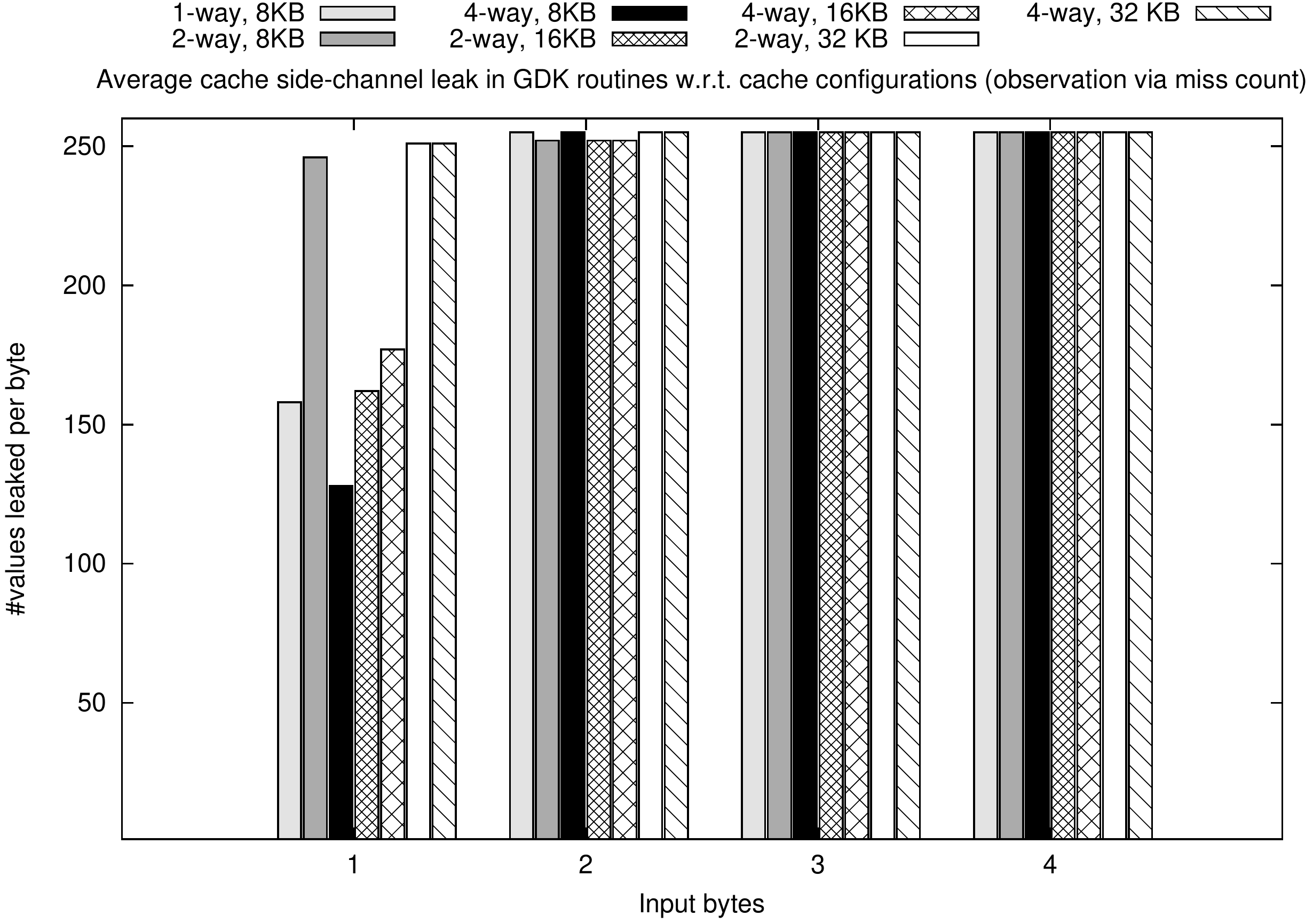}} &
\rotatebox{0}{
\includegraphics[scale = 0.33]{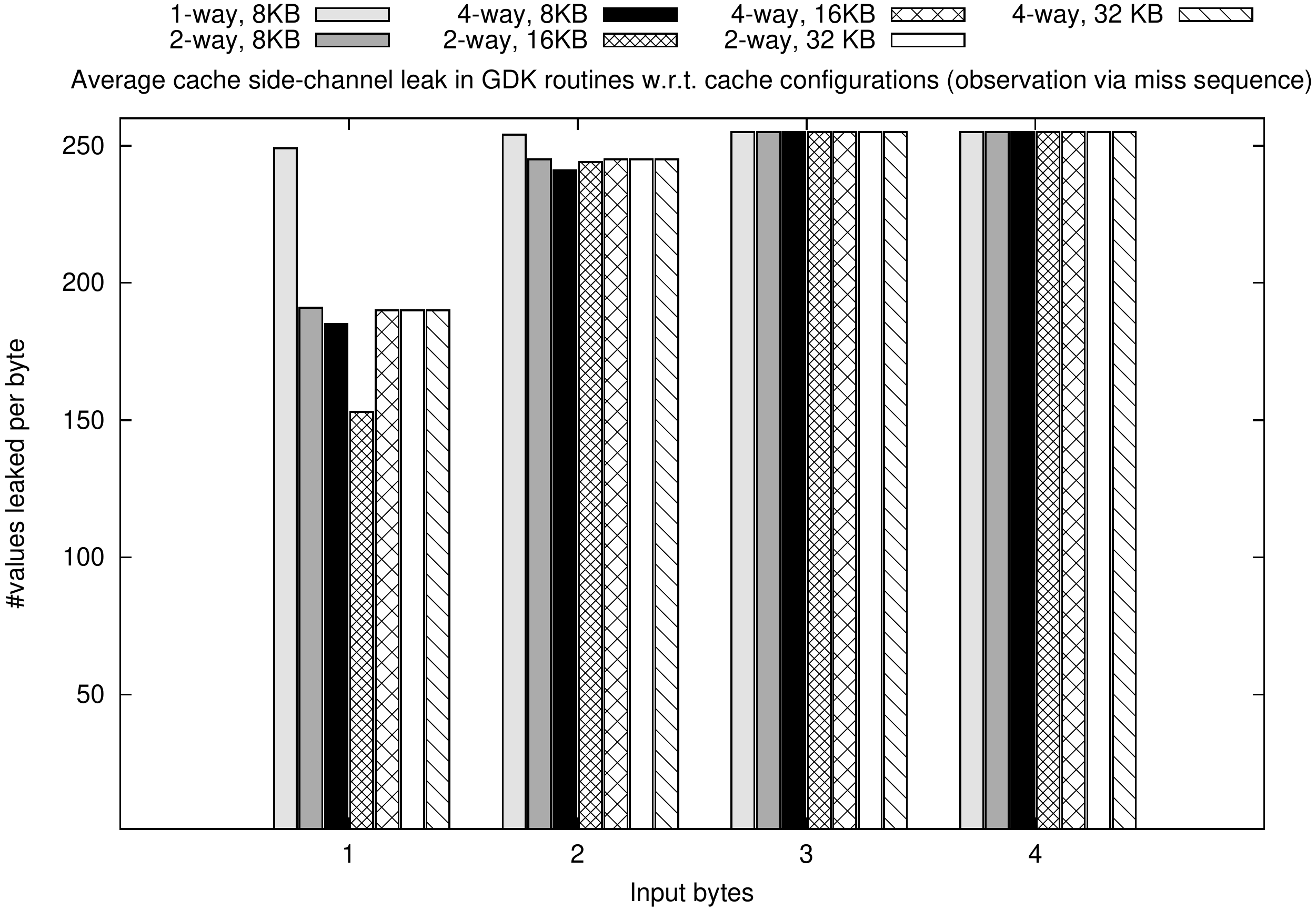}}\\
\textbf{(a)} & \textbf{(b)}
\end{tabular}
\end{center}
\vspace*{-0.1in}
\caption{Information leak in Linux GDK library w.r.t. different observer models}
\label{fig:gdk-result-sens}
\end{figure*}

\subsection{Experience with RC4}
\label{sec:evaluation-rc4}
RC4~\cite{openssl-url} is a stream cipher. It uses variable length key (between 40 
and 2048 bits) and it is considered to be vulnerable in many applications. In our 
evaluation, we studied how RC4 leaks information through cache side channels. We 
analyzed the OpenSSL version of RC4 implementation, where we fixed the size of key 
to be 64 bits. Figures~\ref{fig:rc4-result-sens}(a)-(b) outline our findings. 
Figure~\ref{fig:rc4-result-sens}(a) summarizes our results for miss-count-based 
observer models. \texttt{CHALICE} highlights information being leaked about the 
first byte. For bigger cache sizes ({\em e.g.} $>$ 16KB), such information leak 
disappears, as the executions of RC4 only suffer the minimum number 
of misses to load all the memory blocks into the cache.

Figure~\ref{fig:rc4-result-sens}(b) highlights information leaks when the 
cache hit/miss characteristics of arbitrary memory accesses are observed. 
With respect to such observations, we identify that a substantial information 
may leak (254 values out of a total of 256) about the first byte. However, 
with bigger cache sizes, such information leak disappears. 

\begin{table*}
\caption{$T_1$ captures average time taken for one solver call, 
$T_{byte}$ captures the average time taken to check information leak for one 
input byte and $T_{all}$ captures the time taken to check information leak 
via all predicates in $\mathbf{P_{byte}}$ ({\em cf.} Section~\ref{sec:evaluation-input-predicates})}
\begin{center}
{\scriptsize
 \begin{tabular}{|c|c|c|c|c|c|c|c|c|c|c|}
   \hline
    Subject program & \multicolumn{5}{|c|}{Observation via total miss count} & \multicolumn{5}{|c|}
    {Observation via hit/miss of an arbitrary access}\\
    \cline{2-11}
    & Constraint size & Peak mem. & $T_1$ & $T_{byte}$ & $T_{all}$ & Constraint size & Peak mem. & 
    $T_1$ & $T_{byte}$ & $T_{all}$\\
    \hline
    AES~\cite{aes-impl} & 144072 & 261M & $\approx$ 20 sec & 1 hour & 16 hours & 1580 & 105M & $<$ 1 sec & $\approx$ 1 min  
    &  16 min \\
    \hline
    AES~\cite{openssl-url} & 21444 & 129M & $\approx$ 18 sec & 77 min & 20 hours & 265 & 90M & $<$ 1 sec & $\approx$ 2 min & 45 min\\
	\hline    
    DES~\cite{openssl-url} & 53808 & 127M & $\approx$ 10 sec & 50 min & 8 hours & 1809 & 35M & $<$ 1 sec & $\approx$ 1 min 
    & 12 min \\
    \hline
    RC4~\cite{openssl-url} & 38622 & 1.1G & $\approx$ 4 sec & 15 min & 4 hours & 490 & 32M & $<$ 1 sec & $\approx$ 1 min 
    & 16 min\\
    \hline
    RC5~\cite{openssl-url} & 0 & 28.3M & $\approx$ 15 sec & $\approx$ 15 sec & $\approx$ 15 sec & 0 & 29.2M & $\approx$ 14 sec & $\approx$ 14 sec 
    & $\approx$ 14 sec \\
    \hline
    GDK & 21 & 102M & $<$ 1 sec & $<$ 1 sec & $\approx$ 2 min & 21 & 100M & $<$ 1 sec & $<$ 1 sec & $\approx$ 1 min \\
  	\hline
\end{tabular}}
\end{center}
\vspace*{-0.2in}
\label{tab:time}
\end{table*}

\subsection{Experience with RC5}
\label{sec:evaluation-rc5}
RC5~\cite{openssl-url} is a symmetric block cipher, which is suitable for both 
software and hardware implementation. RC5 has a variable word size and a 
variable-length secret key. In our evaluation, we analyzed the RC5 implementation 
available in the OpenSSL library and we fixed size of both the plaintext and the 
secret key to be 16 bytes. 

During the execution of RC5, our tool \texttt{CHALICE} does not report any symbolic 
memory address. This means, for any memory-related 
instruction, the referenced address is {\em independent} of secret key. As a result, 
cache performance ({\em i.e.} number of cache misses or the sequence of hits and 
misses) of RC5 is unrelated to input and does not exhibit information leak with 
respect to the observer models studied in this paper. It is also worthwhile to mention 
that RC5 does not have any key-dependent branches. Therefore, we can turn the 
report generated by \texttt{CHALICE} into verification, meaning that we can prove 
the absence of information leak according to the observer models explored 
in this paper.

\subsection{Experience with GDK Library}
\label{sec:evaluation-gdk}
Figures~\ref{fig:gdk-result-sens}(a)-(b) present the average information leak 
discovered in routines \texttt{gdk\_keyval\_to\_unicode} and \texttt{gdk\_keyval\_name} 
from the Linux GDK library. As shown in Figure~\ref{fig:gdk-result-sens}, several 
scenarios lead to a complete disclosure of information for the third and the 
fourth byte of input ({\em i.e.} 255 out of 256 values are leaked for 
these bytes). Moreover, the reported information leak persists across a wide-range 
of cache configurations. Upon close inspection, we discovered that the cache 
behaviour of \texttt{gdk\_keyval\_to\_unicode} and \texttt{gdk\_keyval\_name} is 
primarily dominated by the number of cold cache misses, which, in turn is heavily 
influenced by the path executed in the respective routine. As a result, observing 
the cache performance may lead to a disclosure of the (set of) paths taken in 
\texttt{gdk\_keyval\_to\_unicode} and \texttt{gdk\_keyval\_name}. 
Since we include path condition $pc$ within our symbolic cache model $\Gamma (pc)$ 
({\em cf.} Constraint~(\ref{eq:gamma-direct-miss-count}) and 
Constraint~(\ref{eq:gamma-lru-miss-count})), we can accurately quantify 
the information leak even in the presence of multiple program paths.

\subsection{Analysis Time}
\label{sec:analysis-time}

Table~\ref{tab:time} outlines the analysis time for different subject programs 
while using a direct-mapped 8KB cache. 
%
In most cases, a single call to the solver, which reports information leak 
via unsatisfiability check 
({\em e.g.} Constraint~(\ref{eq:heuristic-leak})), is efficient. 
Due to the repeated calls to solver, checking the information 
leakage, for the entire input space, takes significant time. 
However, \texttt{CHALICE} incorporates {\em anytime} strategy, meaning the 
more time it runs the more accurately it can quantify the information leak. 
%
Besides, the timing reported in Table~\ref{tab:time} is either consistent 
or decreases with increasing cache size. This is due to the fact that our 
symbolic cache model uses the notion of cache conflict to encode the cache 
behaviour and the size of our model does not vary dramatically with 
increasing cache size.  
Finally, the performance of \texttt{CHALICE} can be improved drastically 
using a parallel implementation. For instance, we can assign one or more 
independent threads to check information leaked about each input byte.
We plan to implement such a parallel version of \texttt{CHALICE} in the 
future. Table~\ref{tab:time} only reports timing for a sequential implementation 
in this paper.

\subsection{Discussion}
For the sake of brevity, we have only presented the quantification of 
information leak discovered through \texttt{CHALICE}. Of course, due 
to the symbolic nature of our analysis, \texttt{CHALICE} not only quantifies 
information leak, it also highlights which values might leak through 
a potential cache attack. Furthermore, for each memory-related instruction, 
\texttt{CHALICE} highlights the set of input values that may leak for 
a given execution. In summary, the report generated by \texttt{CHALICE} 
can be leveraged for debugging and fixing critical information leak 
scenarios. Some  of the potential debugging strategies would be to 
restructure the code, suppressing or enabling compiler optimizations 
(such as bypassing the cache for certain memory blocks or using 
software-controlled memory) and choosing an appropriate hardware 
platform. In future, we plan to use \texttt{CHALICE} to study the 
impact of such hardware/software transformations on information leak.

\section{Related Work}
\label{sec:related-work}
The closest to our work are approaches based on static 
analysis~\cite{cav12-paper,cacheaudit}. These works quantify 
information leak from the static representation of a program. 
In particular, the information leak is quantified via the 
unique number of observations made by an attacker. As a result, 
these works are incapable to highlight critical information 
leak scenarios when a particular observation leaks substantially 
more information than the rest. \texttt{CHALICE} quantifies 
information leak from execution traces and therefore, it does 
not suffer from the aforementioned limitation. Our work is 
orthogonal to approaches proposed in~\cite{ccs14-paper,usenix16-paper}. 
In particular, our approach targets arbitrary software binaries 
and it is not limited to the verification of constant-time 
cryptographic software. Besides, our approach has a significant 
flavor of testing and debugging, as we highlight information 
leak scenarios directly from execution traces.  
A recent work~\cite{max_smt_side_channel_paper} aims to quantify 
side-channel leakage via symbolic execution and Max-SMT. However, 
this work does not take into account side-channel leaks through 
micro-architectural entities, such as caches.


Over the last few decades, cache-based side-channel attacks have emerged to be 
a prevalent class of security breaches for many systems. A detailed account on 
these attacks can be found in the survey~\cite{side-channel-survey-paper}. 
The observer models used in this paper are based on existing cache 
attacks~\cite{djb:2005cache,trace-attack-paper}. However, we believe that 
\texttt{CHALICE} is generic to incorporate more advanced attack 
scenarios~\cite{cachegames,cache_template_paper,brumley:2009cache}, 
as long as the attacks are expressed via the intuition given in 
Section~\ref{sec:check-information-leak}.

To defend against cache-based side-channel attacks, several countermeasures 
have been proposed over the past few years. Some of these countermeasures 
require hardware support, such as designing new cache 
architecture~\cite{rubylee:rpcache} or compiler support, such as devising 
new instruction-based scheduling~\cite{cacheCM:2013scheduler}. 
More recently, the approach described in~\cite{ndss15-paper-diversity} leverages 
on software diversity at runtime to randomize the cache behavior and hence, 
reducing the probability for a potential cache side-channel attack. Our 
proposal is orthogonal to approaches proposing countermeasures. 
Of course, we believe that our framework can be leveraged as a valuable tool 
to discover potential flaws in countermeasures proposed to mitigate cache 
side channels.

Finally, static cache analysis~\cite{ferdinand-rts-paper} is still an active 
research topic. Compared to static cache analysis, our approach has significant 
flavors of testing and debugging. 
Moreover, our approach can highlight memory accesses that leak significant 
information via side-channels. This can be leveraged to drive security-related 
optimizations.

In summary, we propose a new approach to quantify cache side-channel leakage 
from execution traces and {\em not} from the static representation of the 
program. We demonstrate that such an approach clearly has benefits over 
approaches based on static or logical analysis. This is because \texttt{CHALICE} 
can highlight critical information leak scenarios that are impossible to be 
discovered by competitive static or logical analysis.

\section{Concluding Remarks}
\label{sec:conclusion}

In this paper, we propose a new approach to quantify cache side-channel 
leakage. To the best of our knowledge, \texttt{CHALICE} is the first 
work to categorize input segments with respect to memory performance. 
We illustrate that the mechanism of \texttt{CHALICE} is highly desirable 
for security testing of arbitrary software, specifically, to detect the 
amount of information that can leak through memory performance. Besides 
security testing, \texttt{CHALICE} can also be used to discover bugs while 
writing constant-time cryptographic applications. We demonstrate the usage 
of \texttt{CHALICE} to highlight critical information leak scenarios in 
real-world software -- including applications from OpenSSL and Linux GDK 
libraries.

We believe \texttt{CHALICE} provides a platform to lift the state-of-the-art 
in security testing, in particular, detecting security-related flaws due 
to side channels. As a result, we envision to extend \texttt{CHALICE} for 
side channels other than caches and use it to detect the potential of 
advanced side-channel attacks not investigated in this paper. We hope that 
the core idea of \texttt{CHALICE} would influence regular activities in software 
testing and in testing regressions.

\balance

{
\bibliographystyle{plain}
\bibliography{paper}}

\begin{thebibliography}{10}

\bibitem{aes-impl}
{A}dvanced {E}ncryption {S}tandard {I}mplementation.
\newblock \url{https://github.com/B-Con/crypto-algorithms}.

\bibitem{klee-url}
{KLEE} {LLVM} execution engine.
\newblock \url{https://klee.github.io/}.

\bibitem{openssl-url}
{OpenSSL} {L}ibrary.
\newblock \url{https://github.com/openssl/openssl/tree/master/crypto}.

\bibitem{llvm-url}
{T}he {LLVM} compiler infrastructure.
\newblock \url{http://llvm.org/}.

\bibitem{model-counting}
{UC} {D}avis, {M}athematics. {L}atte integrale.
\newblock \url{https://www.math.ucdavis.edu/~latte/}.

\bibitem{trace-attack-paper}
Onur Ac{\i}i{\c{c}}mez and {\c{C}}etin~Kaya Ko{\c{c}}.
\newblock Trace-driven cache attacks on {AES}.
\newblock In {\em Information and Communications Security}. Springer, 2006.

\bibitem{usenix16-paper}
Jos{\'{e}}~Bacelar Almeida, Manuel Barbosa, Gilles Barthe, Fran{\c{c}}ois
  Dupressoir, and Michael Emmi.
\newblock Verifying constant-time implementations.
\newblock In {\em {USENIX}}, pages 53--70, 2016.

\bibitem{simplescalar-tool}
Todd Austin, Eric Larson, and Dan Ernst.
\newblock {SimpleScalar}: An infrastructure for computer system modeling.
\newblock {\em Computer}, 35(2), 2002.

\bibitem{oracle}
Earl~T Barr, Mark Harman, Phil McMinn, Muzammil Shahbaz, and Shin Yoo.
\newblock The oracle problem in software testing: A survey.
\newblock {\em IEEE transactions on software engineering}, 41(5):507--525,
  2015.

\bibitem{ccs14-paper}
Gilles Barthe, Gustavo Betarte, Juan~Diego Campo, Carlos~Daniel Luna, and David
  Pichardie.
\newblock System-level non-interference for constant-time cryptography.
\newblock In {\em CCS}, pages 1267--1279, 2014.

\bibitem{djb:2005cache}
Daniel~J Bernstein.
\newblock Cache-timing attacks on {AES}, 2005.

\bibitem{brumley:2009cache}
Billy~Bob Brumley and Risto~M Hakala.
\newblock Cache-timing template attacks.
\newblock In {\em ASIACRYPT}. Springer, 2009.

\bibitem{taint-tracking-paper}
James Clause, Wanchun Li, and Alessandro Orso.
\newblock Dytan: a generic dynamic taint analysis framework.
\newblock In {\em ISSTA}. ACM, 2007.

\bibitem{ndss15-paper-diversity}
Stephen Crane, Andrei Homescu, Stefan Brunthaler, Per Larsen, and Michael
  Franz.
\newblock Thwarting cache side-channel attacks through dynamic software
  diversity.
\newblock In {\em NDSS}, 2015.

\bibitem{svf-paper}
John Demme, Robert Martin, Adam Waksman, and Simha Sethumadhavan.
\newblock Side-channel vulnerability factor: {A} metric for measuring
  information leakage.
\newblock In {\em ISCA}, 2012.

\bibitem{cacheaudit}
Goran Doychev, Boris K{\"o}pf, Laurent Mauborgne, and Jan Reineke.
\newblock {CacheAudit}: a tool for the static analysis of cache side channels.
\newblock {\em TISSEC}, 18(1):4, 2015.

\bibitem{side-channel-survey-paper}
Qian Ge, Yuval Yarom, David Cock, and Gernot Heiser.
\newblock A survey of microarchitectural timing attacks and countermeasures on
  contemporary hardware.
\newblock In {\em Cryptology ePrint Archive}, 2016.
\newblock \url{https://eprint.iacr.org/2016/613.pdf/}.

\bibitem{dart-paper}
Patrice Godefroid, Nils Klarlund, and Koushik Sen.
\newblock {DART:} directed automated random testing.
\newblock In {\em PLDI}, 2005.

\bibitem{cache_template_paper}
Daniel Gruss, Raphael Spreitzer, and Stefan Mangard.
\newblock Cache template attacks: Automating attacks on inclusive last-level
  caches.
\newblock In {\em {USENIX} Security}, 2015.

\bibitem{cachegames}
David Gullasch, Endre Bangerter, and Stephan Krenn.
\newblock Cache games--bringing access-based cache attacks on {AES} to
  practice.
\newblock In {\em IEEE Symposium on Security and Privacy}. IEEE, 2011.

\bibitem{cav12-paper}
Boris K{\"o}pf, Laurent Mauborgne, and Mart{\'\i}n Ochoa.
\newblock Automatic quantification of cache side-channels.
\newblock In {\em CAV}. Springer, 2012.

\bibitem{max_smt_side_channel_paper}
Corina~S. Pasareanu, Quoc-Sang Phan, and Pasquale Malacaria.
\newblock Multi-run side-channel analysis using symbolic execution and max-smt.
\newblock In {\em CSF}, 2016.

\bibitem{cacheCM:2013scheduler}
Deian Stefan, Pablo Buiras, Edward~Z Yang, Amit Levy, David Terei, Alejandro
  Russo, and David Mazi{\`e}res.
\newblock Eliminating cache-based timing attacks with instruction-based
  scheduling.
\newblock In {\em ESORICS}, pages 718--735. Springer, 2013.

\bibitem{ferdinand-rts-paper}
Henrik Theiling, Christian Ferdinand, and Reinhard Wilhelm.
\newblock Fast and precise {WCET} prediction by separated cache and path
  analyses.
\newblock {\em Real-Time Systems}, 18(2-3), 2000.

\bibitem{tromer-paper}
Eran Tromer, Dag~Arne Osvik, and Adi Shamir.
\newblock Efficient cache attacks on aes, and countermeasures.
\newblock {\em Journal of Cryptology}, 23(1):37--71, 2010.

\bibitem{rubylee:rpcache}
Zhenghong Wang and Ruby~B. Lee.
\newblock New cache designs for thwarting software cache-based side channel
  attacks.
\newblock In {\em ISCA}, pages 494--505, 2007.

\end{thebibliography}

\end{document}